\documentclass[a4paper,12pt]{article}
\usepackage{graphicx}
\usepackage{hyperref}
\usepackage{amsmath}
\usepackage{amscd,amssymb}
\usepackage{xcolor}

\newtheorem{Resume}[equation]{R\'{e}sum\'{e}}

\newcommand{\cE}{\mathcal{E}}

\newcommand{\cH}{\mathcal{H}}

\newcommand{\fD}{\mathfrak{D}}

\newcommand{\xX}{\textsf{X}}

\newcommand{\id}{{1\!\!1}}

\def\mD{{\mathfrak D}}

\def\mg{{\mathfrak g}}

\def\mg{{\mathfrak g}}

\def\mD{{\mathfrak D}}

\def\mD{{\mathfrak D}}

\def\mg{{\mathfrak g}}

\def\mg{{\mathfrak g}}

\def\cR{{\mathcal R}}

\def\cJ{{\mathcal J}}

\def\cA{{\mathcal A}}

\def\cC{{\mathcal C}}

\def\cH{{\mathcal H}}

\def\cE{{\mathcal E}}

\def\cO{{\mathcal O}}

\def\mD{{\mathfrak D}}

\def\mD{{\mathfrak D}}

\def\mg{{\mathfrak g}}

\def\mg{{\mathfrak g}}

\newtheorem{proposition}{Proposition}[section]

\newtheorem{definition}{Definition}[section]

\newtheorem{remark}{Remark}[section]
\newtheorem{note}{Note}[section]

\newcommand{\beq}{\begin{eqnarray}}
\newcommand{\eeq}{\end{eqnarray}}
\numberwithin{equation}{section}

\begin{document}

\begin{center}
{\large\bf Hilbert Schemes, Verma Modules, and Spectral Functions of Hyperbolic Geometry with  Application
to Quantum Invariants}

\end{center}

\vspace{0.1in}

\begin{center}
{\large
A. A. Bytsenko $^{(a)}$
\footnote{E-mail: aabyts@gmail.com},
M. Chaichian $^{(b)}$
\footnote{E-mail: masud.chaichian@helsinki.fi}
and A. E. Gon\c{c}alves $^{(a)}$
\footnote{E-mail: aedsongoncalves@gmail.com}}

\vspace{5mm}
\vspace{0.2cm}
$^{(a)}$
{\it
Departamento de F\'{\i}sica, Universidade Estadual de
Londrina\\ Caixa Postal 6001,
Londrina-Paran\'a, Brazil}

\vspace{0.2cm}
$^{(b)}$
{\it
Department of Physics, University of Helsinki\\
P.O. Box 64, FI-00014 Helsinki, Finland}

\end{center}

\vspace{0.1in}

\begin{abstract}
In this article we exploit Ruelle-type spectral functions and analyze the Verma module over Virasoro algebra,
boson-fermion correspondence, the analytic torsion, the Chern-Simons and $\eta$ invariants, as well as
the generation function associated to dimensions of the Hochschild homology of the crossed product
$\mathbb{C}[S_n]\ltimes \cA^{\otimes n}$ ($\cA$ is the $q$-Weyl algebra).
After analysing the Chern-Simons and $\eta$ invariants of Dirac operators by
using irreducible $SU(n)$-flat connections on locally symmetric manifolds of
non-positive section curvature, we describe the exponential action for the
Chern-Simons theory.
\end{abstract}

\vspace{0.1in}

\begin{flushleft}
MSC \, 05A30 (q-Calculus and related topics)


\vspace{0.3in}
March 2019
\end{flushleft}

\newpage

\tableofcontents


\section{Introduction}

{\bf Finite-dimensional Lie algebras and certain applications.}
Originally the Lie theory has been viewed as a Lie group and a group of symmetries of an
algebraic or a geometric object. In addition, the initial role of the corresponding Lie algebra
was in classifying symmetries of ordinary differential equations, and it has become essential in describing the symmetries of many physical systems.
Finite-dimensional Lie algebras can frequently be involved in the definition for an infinite-dimensional case.
In this connection, one can mention the finite-dimensional semi-simple Lie algebras and their
representations, which have been classified.
The characters of the irreducible representations of $gl(n)$ can be given by (composite) symmetric functions.
Lie superalgebras (that is ${\mathbb Z}_2$ graded Lie algebras with a graded commutator) were introduced
and classified in \cite{Kac}.

Lie algebras and superalgebras have useful realizations
in terms of bosonic and fermionic creation and annihilation operators, which promotes to description the kinematical
or dynamical symmetries of many physical systems (for example, supersymmetric quantum mechanics).
Another type of algebras, which has a wide variety of applications in physics is the so-called quantum groups.
These algebras may be regarded as a deformation, depending on a parameter $q$, of the universal enveloping
algebras of a semi-simple Lie algebras. Thus they are not finite-dimensional algebras, but are finitely generated.

On the time being quantum groups can be considered as an example of quasi-triangular Hopf algebras.
For each quantum group there exists
an universal $R$-matrix which intertwines with the action of the coproduct.
Various realizations of quantum (super)groups have been given in terms of $q$-deformed bosonic and fermion oscillators.
These realizations have found use in quantum-mechanical applications.

{\bf Certain applications of infinite-dimensional Lie algebras.}
A particular class of infinite-dimensional algebras, (affine) Kac-Moody algebras (which were introduced in the late 1960's)
plays a part in many diverse areas of mathematics and physics.
Their finite-dimensional counterparts, as well as the representation theory, has been developed and various realizations
have been constructed. The simple Lie algebras can be realized in terms of a finite number of fermionic/bosonic modes;
simple Kac-Moody algebras have various vertex operator realizations -- that is, in terms of a finite number of bosonic
free fields, the modes of which, generate the Heisenberg algebra.
All simple (twisted and untwisted) Kac-Moody algebras can be embedded in the
infinite-dimensional algebra $gl(\infty)$ of infinite matrices with a finite number of non-zero
entries, which has a simple realization in terms of generators of the Clifford algebra.

Another type of infinite-dimensional algebra which arises in different areas of physics is the Virasoro algebra.
This algebra is the algebra of conformal transformations in two-dimensions.
It has been shown that the operator algebra structure of two-dimensional
conformally-invariant quantum field theories is determined by the representation theory of the Virasoro algebra.

Thus the complete classification of the unitary, irreducible highest weight representations of the Virasoro algebra
became an important problem. Also the two supersymmetric extensions of the Virasoro algebra were proposed.

Finally, we note the quantum affine algebras which are $q$-deformations of Kac-Moody algebras $\widehat{g}$.
One of the main application of quantum affine algebras has been to study the degeneracies in the spectrum of the
(anti-ferromagnetic) XXZ quantum spin chain Hamiltonian in the thermodynamic limit \cite{Davies}.
(In this circuit the Hamiltonian of the spin 1/2 XXZ chain is related to the derivation operator of
$U_q (\widehat{sl(2)})$, and the space of states, realized by the infinite tensor product of two-dimensional vector spaces,
is isomorphic to the tensor product of the certain irreducible $U_q (\widehat{sl(2)})$ modules.)

\subsection{Structure of the article and our key results}

{\bf Preliminary observation and explanation.}
In this paper we shall make considerable use of Ruelle-type functions $\cR(s)$.
$\cR(s)$ should give a well-balanced description of the content of the whole article.
These functions are connected to symmetric functions (so-called S-functions $s_\lambda(x)$), which play important role
in the representation theory of finite-dimensional classical Lie algebras. As an example note the character
of the irreducible representation of $gl(n)$ labelled by a standard partition $\lambda$ -- the S-function
$s_\lambda(x)$.

{\bf The organization of the article is as follows}: We begin in Sects. \ref{Laplacian} and \ref{Three-geometry}
with Laplasian on forms and definition of Ruelle functions. Then the theory of Ruelle-type functions
(which are an alternating product of more complicate factors, each of them is the Patterson-Selberg zeta function)
with its connection to Euler series is developed in Sect. \ref{Ruelle}.

We discuss the Hilbert scheme of points on surface in Sect. \ref{Scheme}. Explicit formulas for the
dimension of $q$-space of Hirota polynomials, character of the Heisenberg algebra, as well as for the
super-Heisenberg algebra, are deduced in terms of Ruelle-type spectral functions.

We analyze the Verma module over Virasoro algebra in Sect. \ref{Virasoro}. The representation theory of
Virasoro algebras is, in fact, very similar to those for Kac-Moody algebras.
We show that the analytic structure of the formal character of the $Vir$-module ${M(c, h)}$
is determined by the pole structure of the Ruelle function.

In Sect. \ref{cyclic}, we investigate the analytic torsion $\tau^{\bf an}$ and the $\eta$-invariant via cyclic homology.
By using the irreducible $SU(n)$-flat connections on a locally symmetric manifolds of non-positive section
curvature, we analyze the Chern-Simons and $\eta(s, {\mathfrak D})$ invariants (Sect. \ref{Chern-Simons}) -- topological
invariants of a pair $(X, \rho)$, where $\rho$ is a representation of $\pi_1(X)$. We describe the exponential
action of the Chern-Simons invariant in terms of the Dirac (and twisted Dirac) operators.

In Sect. \ref{Crossed}, we describe the deformation quantization and the crossed products $\mathbb{C}[W]\ltimes \cA^{\otimes n}$,
where $\cA$ is the $q$-Weyl algebra (or any its degeneration); the Weyl group is type $A_{n-1}$ or $B_n$.
The generating function for ${\rm dim}\, HH^i(\mathbb{C}[W]\ltimes \cA^{\otimes n})$ as well as the G\"{o}ttsche
formula are derived in terms of Ruelle-type spectral functions.

\section{Laplacian on forms and Ruelle-type functions}
\label{Laplacian}

Let $L_p$ be a self-adjoint Laplacian on $p$-forms. The heat kernel expansion for this Laplacian,
acting on a compact manifold, is given by:
\begin{proposition}
There exist $\varepsilon,\delta >0$
such that for $0<t<\delta$ the heat kernel expansion for
Laplace operators on a compact manifold $X$ is given by
\begin{equation}
{\rm Tr}\left(e^{-tL_p}\right)= \sum_{0\leq \ell\leq \ell_0} a_\ell
(L_p)t^{-\ell}+ {O}(t^\varepsilon).
\end{equation}
The coefficients $a_\ell(L_p)$ are called Hadamard-Minakshisundaram-De Witt-Seeley coefficients
(or, sometimes, heat kernel, or just heat coefficients)\,.
\end{proposition}

We consider now a simplified version of the Hodge -- de Rham theory.
Let $A$ be a module over $g$; in other words $g(ab)= (ga)b + a(gb)$ for all $g\in \mg$,
$a,b \in A$. Assume also that the considered Lie algebra posssesses a grading, i.e. $\mg$
is a direct sum of its subspaces $\mg_{(\lambda)}$, where $\lambda$ are integers (real or
complex numbers). We assume that $[\mg_{(\lambda)}, \mg_{(\mu)}]\subset \mg_{(\lambda+\mu)}$,
and suppose further that the $\mg$-module $A$ is also graded by homogeneous components $A_{(\mu)}$,
$\mg_{(\lambda)}A_{(\mu)}\subset A_{(\lambda+\mu)}$.

Suppose that the Lie algebra $\mg$ is finite-dimensional or assume that it is graded in such a way that
all the spaces $C_{(\lambda)}^q(\mg)$ are finite dimensional. Thus gradings arise in chain and
cochain spaces, indeed:
\begin{equation}
C_{(\lambda)}^q(\mg; A) = \{c\in C^q(\mg; A)\vert c(g_1,\ldots,g_q) \in A_{(\lambda_1+\ldots+\lambda_q-\lambda)}),
\, g_i\in \mg_{(\lambda_i)}\}.
\end{equation}
$C_q^{(\lambda)}$ is generated by the chains $a\otimes (g_1\wedge \ldots \wedge g_q)$ with $a\in A_{(\mu)}$,
$g_i\in \mg_{(\lambda_i)}$, $\lambda_1+\ldots +\lambda_q+\mu = \lambda$.
In addition $d(C_{(\lambda)}^q(\mg; A)) \subset C^{q+1}_{(\lambda)}(\mg; A)$; $\partial(C_q^{(\lambda)}(\mg; A))
\subset C_{q-1}^{(\lambda)}(\mg; A)$, thus both homology and cohomology asquire gradings. In particular
$H_{(\lambda)}^p(\mg)H_{(\mu)}^q(\mg)\subset H_{(\lambda+\mu)}^{p+q}(\mg)$.
\begin{proposition}
 Each element of the space $H_{(\lambda)}^q(\mg)$ can be represented by a unique harmonic cocycle from
 $C_{(\lambda)}^q(\mg)$; hence there is a natural isomorphism ${\rm ker} L_{p,(\lambda)}^q= H_{(\lambda)}^q(\mg)$.
\end{proposition}

\subsection{The Ray-Singer norm}

Let $\rho$ be an orthogonal representation of $\pi_1(X)$. Using
the Hodge decomposition, the vector space $H(X;\rho)$ of twisted
cohomology classes can be embedded into $\Omega(X;\rho)$ as the
space of harmonic forms. This embedding induces a norm
$|\cdot|^{RS}$ on the determinant line ${\rm det}H(M;\rho)$. The
Ray-Singer norm $||\cdot||^{RS}$ on ${\rm det}H(X;\rho)$ is
defined by \cite{RS}
\begin{equation}
||\cdot||^{RS}\stackrel{def}=|\cdot|^{RS}\prod_{p=0}^{{\rm dim}\,X}
\left[\exp\left(-\frac{d}{ds}
\zeta (s|L_p)|_{s=0}\right)\right]^{(-1)^pp/2}
\mbox{,}
\end{equation}
where the zeta function $\zeta (s|L_p)$ of the Laplacian
acting on the space of
$p$-forms orthogonal to the harmonic forms has been used. For a
closed connected orientable smooth manifold of odd dimension
and for Euler structure
$\eta\in {\rm Eul}(X)$, the Ray-Singer norm of its cohomological
torsion $\tau^{\bf an}(X;\eta)=\tau^{\bf an}(X)\in {\rm det}H(X;\rho)$ is
equal to the positive
square root of the absolute value of the monodromy of $\rho$
along the characteristic class $c(\eta)\in H^1(X)$:
$||\tau^{\bf an}(X)||^{RS}=|{\rm det}_{\rho}c(\eta)|^{1/2}$.
In the special
case where the flat bundle $\rho$ is acyclic, we have
\begin{equation}
\left[\tau^{\bf an}(X)\right]^2
=|{\rm det}_{\rho}c(\eta)|
\prod_{p=0}^{{\rm dim}\,X}\left[\exp\left(-\frac{d}{ds}
\zeta (s|L_p)|_{s=0}\right)\right]^{(-1)^{p+1}p}
\mbox{.}
\label{RS}
\end{equation}
For a  closed oriented hyperbolic three-manifolds of the form
$X = {H}^3/\Gamma$, and for acyclic
$\rho$, the $L^2$-analytic torsion has the form
\cite{Fried,Bytsenko3,Bytsenko4}:
$[\tau^{\bf an}(X)]^2={\mathcal R}(0)$, where
${\mathcal R}(s)$ is the Ruelle function
which can be continued meromorphically to the entire complex plane
$\mathbb C$.
The function ${\mathcal R}(s)$ is an alternating
product of more complicate factors, each of which is a Selberg zeta function $Z(s)$.
The relation between the Ruelle and Selberg zeta functions is:
\begin{equation}
{\mathcal R}(s)=\prod_{n=0}^{{\rm dim}\,X-1}Z
(n+s)^{(-1)^n}\,.
\label{Ruelle1}
\end{equation}
The Ruelle zeta function associated with closed oriented
hyperbolic three-manifold $X$ has the form:
${\mathcal R}(s)=Z(s)Z(2+s)/Z(1+s)$.

\section{Hyperbolic groups}
\label{Hyperbolic}

Let ${\mathfrak a}_0, {\mathfrak n}_0$ denote the Lie algebras of $A, N$ in an
Iwasawa decomposition, $G=KAN$. Since we are interested in
hyperbolic geometry, let us consider the case $G=SO_1(2n,1)$,
$K=SO(2n)$. The complexified Lie algebra ${\mathfrak g}={\mathfrak
g}^{\mathbb C}_0 ={\mathfrak so}(2n+1,{\mathbb C})$ of $G$ is of
the Cartan type $B_n$:
\begin{eqnarray}
 & Q & =  \Big\{\sum_ik_i\varepsilon_i \in {\mathbb R}^n \mid
k_i\in {\mathbb Z}\Big\}\,,
\nonumber \\
& Q^\vee & = \Big\{\sum_ik_i\varepsilon_i\in {\mathbb
R}^n\mid k_i\in {\mathbb Z},\, \sum_ik_i\in 2{\mathbb Z}\Big\},
\nonumber \\
& \triangle & =  \{\pm\varepsilon_i\pm\varepsilon_j,
\pm\varepsilon_i\},
\nonumber \\
& \Pi & =  \{\alpha_1 = \varepsilon_1-\varepsilon_2, \cdots ,
\alpha_{n-1} = \varepsilon_{n-1} -\varepsilon_n, \alpha_n =
\varepsilon_n\},
\nonumber \\
& W & = \{{\rm all}\,\,\,\,{\rm permutations}\,\,\,\,{\rm
and}\,\,\,\,{\rm sign}\,\,\,\,{\rm changes} \,\,\,\,{\rm
of}\,\,\,\,{\rm the}\,\,\,\, \varepsilon_i\} = Aut\,Q\,.
\end{eqnarray}
Here $\Pi$ is a basis of $Q$ over $\mathbb Z$, the matrix
$2\langle\alpha_i,\alpha_j\rangle/\langle\alpha_i, \alpha_i\rangle$ is the Cartan matrix
of the corresponding type. Since the rank of $G$ is one, $\dim
{\mathfrak a}_0=1$ by definition, say
\begin{equation}
{\mathfrak a}_0={\mathbb R}\,H_0 \,\,\,\, {\rm for}\,\,\,\,{\rm
a}\,\,\,\, {\rm suitable} \,\,\,\,{\rm basis}\,\,\,\, {\rm
vector}\,\,\,\, H_0 := {\rm antidiag}(1, \cdots, 1)
\end{equation}
is a $(n+1)\times(n+1)$ matrix. By this choice we have the
normalization $\beta(H_0)=1$, where $\beta: {\mathfrak
a}_0\rightarrow{\mathbb R}$ is the positive root which defines
${\mathfrak n}_0$. Note that the Killing form $\langle\cdot ,\cdot \rangle$ is given by
$(x,y)=(n-1)\,{\rm trace}(xy)$ for $x,\;y\in {\mathfrak g}_0$.
The standard systems of positive roots $\triangle^{+},\triangle^{+}_s$
for ${\mathfrak g}$ and ${\mathfrak k}={\mathfrak k}^ {\mathbb C}_0$ -- the
complexified Lie algebra of $K$, with respect to a Cartan subgroup
$H$ of $G$,\, $H\subset K$, are given by
\begin{equation}
\triangle^{+}=\{\varepsilon_i|1\leq i\leq n\}\cup
\triangle^{+}_s\,, \,\,\,\,\,\,\,
\triangle^{+}_s=\{\varepsilon_i\pm \varepsilon_j|1\leq i<j\leq
n\},
\end{equation}
and
\begin{equation}
\triangle^{+}_n \stackrel{\rm def}{=}\{\varepsilon_i|1\leq i\leq
n\}
\end{equation}
is the set of positive non-compact roots. Here,
\begin{equation}
\langle\varepsilon_i , \varepsilon_j\rangle = \frac{\delta_{ij}}{\langle H_0,
H_0\rangle} = \frac{\delta_{ij}}{2(2n-1)}\,,\,\,\,\,\,\,\,
(\varepsilon_i \pm \varepsilon_j , \varepsilon_i \pm
\varepsilon_j) = \frac{1}{2n+1}\,\,\,,\,\,\, i<j\,,
\end{equation}
i.e. $ \langle\alpha , \alpha\rangle = (2n-1)^{-1},\,\, \forall \alpha\in
\triangle^{+}_n.\, $ Let $\tau = \tau^{(j)}$ be a representation of
$K$ on $\Lambda^j{\mathbb C}^{2n}$. The highest weight of $\tau$,
$\Lambda_{\tau^{(j)}}=\Lambda_j$, is
\begin{equation}
\left\{ \begin{array}{ll} \varepsilon_1 + \cdots +
\varepsilon_j,\,\,\,\,\,\,
&{\rm if}\,\,\, j\leq n,\\
\varepsilon_1 + \cdots + \varepsilon_{2n-j},\,\,\,\,\,\, &{\rm
if}\,\,\, j>n.
\end{array} \right.
\end{equation}
Writing $ \langle\Lambda_j, \Lambda_j+2\rho_n\rangle = \langle\Lambda_j, \Lambda_j\rangle
+ \langle\Lambda_j, 2\rho_n\rangle, \,\, \rho_n = \sum_{i=1}^n
(n-i)\varepsilon_i $, for $j\leq n$ we have
\begin{eqnarray}
\langle\Lambda_j , \Lambda_j\rangle & = & \langle\sum_{p=1}^j\varepsilon_p ,
\sum_{q=1}^j\varepsilon_q\rangle = \sum_{p, q =1}^j \langle\varepsilon_p ,
\varepsilon_q\rangle =\sum_{p=1}^j \langle\varepsilon_p , \varepsilon_p\rangle =
\frac{j}{\langle H_0 , H_0\rangle}\,,
\\
\langle\Lambda_j ,  2\rho_n\rangle & = & \langle\sum_{p=1}^j\varepsilon_p ,
2\sum_{i=1}^j(n-i)\varepsilon_i\rangle +
2\sum_{i=j+1}^n(n-i)\varepsilon_i = 2\sum_{p=1}^j\langle\varepsilon_p ,
(n-p)\varepsilon_p\rangle
\nonumber \\
& = & \frac{2nj}{\langle H_0 , H_0\rangle} -2\sum_{p=1}^jp\langle\varepsilon_p ,
\varepsilon_p\rangle = \frac{2nj}{\langle H_0 , H_0\rangle} -\frac{j(j+1)}{\langle H_0 ,
H_0\rangle^2}.
\end{eqnarray}
Therefore,
\begin{equation}
\langle\Lambda_j , \Lambda_j+2\rho_n\rangle = \frac{j(2n+1)}{\langle H_0 , H_0\rangle} -
\frac{j(j+1)}{\langle H_0 , H_0\rangle^2}.
\end{equation}
In the case $j>n$, we have
\begin{eqnarray}
\langle\Lambda_j , \Lambda_j\rangle & = & \langle\sum_{p=1}^{2n-j}\varepsilon_p ,
\sum_{q=1}^{2n-j}\varepsilon_q\rangle = \sum_{p=1}^{2n-j}\langle\varepsilon_p
, \varepsilon_p\rangle =\frac{2n-j}{\langle H_0 , H_0\rangle},
\\
\langle\Lambda_j , 2\rho_n\rangle & = & 2\langle\sum_{p=1}^{2n-j}\varepsilon_p ,
\sum_{i=1}^{n}(n-i)\varepsilon_i\rangle
\nonumber \\
& = & 2\langle\sum_{p=1}^{2n-j}\varepsilon_p ,
\sum_{i=1}^{2n-j}(n-i)\varepsilon_i\rangle +
\sum_{i=2n-j+1}^{n}(n-i)\varepsilon_i
\nonumber \\
& = & 2\langle\sum_{p=1}^{2n-j}\varepsilon_p ,
\sum_{i=1}^{2n-j}n\varepsilon_i\rangle - \sum_{i=1}^{2n-j}i\varepsilon_i
=\frac{2n(2n-j)}{\langle H_0 , H_0\rangle} - 2\sum_{i=1}^{2n-j}i(\varepsilon_i
, \varepsilon_i)
\nonumber \\
& = & \frac{2n(2n-j) - (2n-j)(2n-j+1)}{\langle H_0 , H_0\rangle} =
\frac{(2n-j)(j-1)}{\langle H_0 , H_0\rangle}.
\end{eqnarray}

\subsection{Highest-weight modules}
\label{modules}

For $\Lambda_j = \triangle_{s}^{+}-$ highest weight of
$K=SO(2n)$ on $\Lambda^j{\mathbb C}^{2n}$, we have
\begin{equation}
\langle\Lambda_j , \Lambda_j\rangle + 2\delta_n = \frac{2nj-j^2}{\langle H_0 , H_0\rangle}
= \frac{2nj-j^2}{2(2n-1)} \,\,\,\,\,\,\, {\rm for}\,\,\,\, 0\leq
j\leq 2n.
\end{equation}
Let ${\mathfrak h}_0$ be the Lie algebra of $H$ and let
${\mathfrak h}^{*}_{\mathbb R}={\rm Hom}(\sqrt{-1}{\mathfrak h}_0,
{\mathbb R})$ be the dual space of the real vector space
$\sqrt{-1}{\mathfrak h}_0$. Thus, the $\{\varepsilon_i\}_{i=1}$
are an ${\mathbb R}$-basis of ${\mathfrak h}^{*}_{\mathbb R}$. Of
interest are the {\em integral} elements $f$ of ${\mathfrak
h}^{*}_{\mathbb R}$:
\begin{eqnarray}
f\stackrel{def}{=}\{\lambda\in {\mathfrak h}^{*}_{\mathbb R}\mid
\langle \lambda\mid\alpha\rangle\equiv \frac{2\langle\lambda,\alpha\rangle}
{\langle\alpha,\alpha\rangle}\in {\mathbb Z},\,\,\, \forall \alpha\in
\triangle^{+}\}.
\end{eqnarray}
Then we have
\begin{equation}
\langle \lambda\mid\varepsilon_i\rangle =  2\lambda_i\,\,\,\,\,
{\rm for}\,\,\,\,\, 1\leq i\leq n, \,\,\,\,\,\,\, \langle
\lambda\mid\varepsilon_i\pm\varepsilon_j\rangle
= \lambda_i\pm \lambda_j \,\,\,\,\,\,  {\rm for}\,\,\,\,\,1\leq
i<j\leq n,
\end{equation}
where we shall write $\lambda =\sum_{j=1}^n\lambda_j\varepsilon_j$
for $\lambda\in {\mathfrak h}^{*}_{\mathbb R},\, \lambda_j\in
{\mathbb R}$. Then clearly
\begin{eqnarray}
f & = & \{\lambda\in {\mathfrak h}^{*}_{\mathbb R}\mid
2\lambda_i\,\in \,{\mathbb Z}\}\,\,\,\,\,\,\,\,\,\,\,\,\,\, {\rm
for}\,\,\,\, 1\leq i\leq n,
\\
f & = & \{\lambda\in {\mathfrak h}^{*}_{\mathbb R}\mid
\lambda_i\pm \lambda_j \in \,{\mathbb Z}\}\,\,\,\,\,{\rm
for}\,\,\,\,1\leq i<j\leq n.
\end{eqnarray}
Let $ \rho_s= (1/2)\sum_{\alpha\in\triangle^{+}_s}\alpha,\,
\rho_n= (1/2)\sum_{\alpha\in\triangle^{+}_n}\alpha,\, \rho
=\rho_s+\rho_n= (1/2)\sum_{\alpha\in\triangle^{+}}\alpha. $ Then
\begin{equation}
\rho_s=\sum_{i=1}^n(n-i)\varepsilon_i,\,\, \rho_n=
\frac{1}{2}\sum_{i=1}^n\varepsilon_i,\,\, \rho =\sum_{i=1}^n(n-i-
\frac{1}{2})\varepsilon_i
\end{equation}
are all integral. The elements $\lambda$ of $f$ correspond to
characters $e^{\lambda}$ of $H$. We can deduce the specialization
formula \cite{Kac_book}:
\begin{equation}
\prod_{j\geq 1}(1-q^j)^{{\rm dim}\, {\mathfrak g}_j(s)} = \sum_{w\in
W^s}\varepsilon (w)K_s(w(\rho))q^{\langle\rho - w(\rho),
h^s\rangle}\,.
\end{equation}
Here,
\begin{equation}
K_s(\lambda) = \prod_{\alpha\in \triangle_{s}^{+}} \langle
\lambda, \alpha^{\vee}\rangle/\langle \rho_s,
\alpha^{\vee}\rangle,\,\,\,\,\,\,\, \triangle_{s}^{+} =
\{\alpha\in \triangle_+ \vert \langle\alpha, h^s\rangle = 0 \}
\end{equation}
and $\rho_s$ is the half-sum of roots from $\triangle_{s+}$; $W^s$
is a system of representatives of left cosets of the subgroup
$W_s$ generated by $r_\alpha$, $\alpha \in \triangle_{s+}$ in $W$,
so that $W= W_sW^s$; ${\mathfrak g}(A) = \oplus_j {\mathfrak
g}_j(s)$ is the $\mathbb Z$-gradation of ${\mathfrak g}(A)$ of
type $s$. For the case $G= SO_1(2n, 1), K=SO(2n)$ we have
\begin{equation}
K_s(\lambda) = \prod_{\alpha\in \triangle_{s}^{+}}\langle\lambda,
\alpha\rangle/\langle\rho_s, \alpha\rangle = \prod_{1\leq i<j\leq n}
\frac{\lambda_i^2 -\lambda_j^2}{(2n-i-j)(j-i)}\,.
\end{equation}

In general,generating functions adopt the form of expressions for the
Euler series $ \Pi_n(1-q^n)^{{\rm dim}\,{\mathfrak
g}_n}, $ $ \Pi_n(1-q^n)^{{\rm rank}\,{\mathfrak g}_n}\,. $ These
formulas are associated with the dimensions of the homology of
appropriate topological spaces. Formally the product expansion
can be written as follows:
\begin{eqnarray}
\prod_n(1-q^n)^{{\rm dim}\,{\mathfrak g}_{n}} & = &
\sum_{m,\lambda}(-1)^m q^\lambda {\rm
dim}\,H_m^{(\lambda)}({\mathfrak g})\,\, = \,\,
\sum_{\lambda}q^\lambda {K}^{(\lambda)}({\mathfrak g}),
\\
{K}^{(\lambda)}({\mathfrak g})  & = & \sum_{m}(-1)^m {\rm
dim}\,H_m^{(\lambda)}({\mathfrak g}).
\end{eqnarray}

\subsection{Spectral functions of hyperbolic three-geometry}
\label{Three-geometry}

Now let us consider three-geometry with an orbifold description $H^3/\Gamma$. The complex unimodular group $G=SL(2, {\mathbb C})$
acts on the real hyperbolic three-space $H^3$ in a standard way, namely for $(x,y,z)\in H^3$ and $g\in G$, one gets
$g\cdot(x,y,z)= (u,v,w)\in H^3$. Thus for $r=x+iy$,\,
$g= \left[ \begin{array}{cc} a & b \\ c & d \end{array} \right]$,
$
u+iv = [(ar+b)\overline{(cr+d)}+ a\overline{c}z^2]\cdot
[|cr+d|^2 + |c|^2z^2]^{-1},\,
w = z\cdot[
{|cr+d|^2 + |c|^2z^2}]^{-1}\,.
$
Here the bar denotes the complex conjugation. Let $\Gamma \in G$ be the discrete group of $G$
defined as
\begin{eqnarray}
\Gamma & = & \{{\rm diag}(e^{2n\pi ({\rm Im}\,\tau + i{\rm Re}\,\tau)},\,\,  e^{-2n\pi ({\rm Im}\,\tau + i{\rm Re}\,\tau)}):
n\in {\mathbb Z}\}
= \{{\mathfrak g}^n:\, n\in {\mathbb Z}\}\,,
\nonumber \\
{\mathfrak g} & = &
{\rm diag}(e^{2\pi ({\rm Im}\,\tau + i{\rm Re}\,\tau)},\,\,  e^{-2\pi ({\rm Im}\,\tau + i{\rm Re}\,\tau)})\,.
\end{eqnarray}
One can define a Selberg-type zeta function for the group
$\Gamma = \{{\mathfrak g}^n : n \in {\mathbb Z}\}$ generated by a single hyperbolic element of the form
${\mathfrak g} = {\rm diag}(e^z, e^{-z})$, where $z=\alpha+i\beta$ for $\alpha,\beta >0$. In fact, we will take
$\alpha = 2\pi {\rm Im}\,\tau$, $\beta= 2\pi {\rm Re}\,\tau$. For the standard action of $SL(2, {\mathbb C})$ on $H^3$ one has
\begin{equation}
{\mathfrak g}
\left[ \begin{array}{c} x \\ y\\ z \end{array} \right]
=
\left[\begin{array}{ccc} e^{\alpha} & 0 & 0\\ 0 & e^{\alpha} & 0\\ 0
& 0 & \,\,e^{\alpha} \end{array} \right]
\left[\begin{array}{ccc} \cos(\beta) & -\sin (\beta) & 0\\
\sin (\beta) & \,\,\,\,\cos (\beta) & 0
\\ 0 & 0 & 1 \end{array} \right]
\left[\begin{array}{c} x \\ y\\ z \end{array} \right]
\,.
\end{equation}
Therefore, ${\mathfrak g}$ is the composition of a rotation in ${\mathbb R}^2$ with complex eigenvalues
$\exp (\pm i\beta)$ and a dilatation $\exp (\alpha)$. There exists the Patterson-Selberg spectral function $Z_\Gamma (s)$,
meromorphic on $\mathbb C$. The Patterson-Selberg function can be attached to ${H}^3/\Gamma$ as follows \cite{Perry,PW}:
\begin{equation}
Z_\Gamma(s) :=\prod_{k_1, k_2 \in \mathbb{Z}_+ \cup \{0\}} [1-(e^{i\beta})^{k_1}(e^{-i\beta})^{k_2}e^{-(k_1+k_2+s)\alpha}]\,.
\label{zeta00}
\end{equation}
Zeros of $Z_\Gamma (s)$ are the complex numbers
$
\zeta_{n,k_{1},k_{2}} = -\left(k_{1}+k_{2}\right)+i\left(k_{1}-
k_{2}\right)\beta/\alpha+ 2\pi  in/\alpha\,\,\,
(n \in {\mathbb Z}).
$

\subsection{Euler series and Ruelle-type spectral functions}
\label{Ruelle}

The most important Euler series can be represent in the form of the
Ruelle-type (Patterson-Selberg) spectral function $\cR(s)$ of hyperbolic three-geometry.
\begin{eqnarray}
\prod_{n=\ell}^{\infty}(1- q^{an+\varepsilon})
& = & \prod_{p=0, 1}Z_{\Gamma}(\underbrace{(a\ell+\varepsilon)(1-i\varrho(\tau))
+ 1 -a}_s + a(1 + i\varrho(\tau)p)^{(-1)^p}
\nonumber \\
& = &
\cR(s = (a\ell + \varepsilon)(1-i\varrho(\tau)) + 1-a),
\label{R1}
\\
\prod_{n=\ell}^{\infty}(1+ q^{an+\varepsilon})
& = &
\prod_{p=0, 1}Z_{\Gamma}(\underbrace{(a\ell+\varepsilon)(1-i\varrho(\tau)) + 1-a +
i\sigma(\tau)}_s
+ a(1+ i\varrho(\tau)p)^{(-1)^p}
\nonumber \\
& = &
\cR(s = (a\ell + \varepsilon)(1-i\varrho(\tau)) + 1-a + i\sigma(\tau))\,.
\label{R2}
\end{eqnarray}
Here $q= \exp(2\pi i\tau)$, $\varrho(\tau) =
{\rm Re}\,\tau/{\rm Im}\,\tau$,
$\sigma(\tau) = (2\,{\rm Im}\,\tau)^{-1}$,
$a$ is a real number, $\varepsilon, b\in {\mathbb C}$, $\ell \in {\mathbb Z}_+$.

Next let us introduce some well-known functions and their modular properties under the action of
$SL(2, {\mathbb Z})$. The special cases associated with (\ref{R1}), (\ref{R2}) are (see  \cite{Kac_book}):
\begin{eqnarray}
\varphi_1(q) & = & q^{-\frac{1}{48}}\prod_{m=1}^\infty
(1-q^{m+\frac{1}{2}})\,\, = q^{-\frac{1}{48}}\cR(s= (1+1/2)(1-i\varrho(\tau)))\,\, \nonumber \\
& = & \frac{\eta_D(q^{\frac{1}{2}})}{\eta_D(q)}\,,
\\
\varphi_2(q) & = & q^{-\frac{1}{48}}\prod_{m=1}^\infty
(1+q^{m+\frac{1}{2}})\,\, = q^{-\frac{1}{48}}\cR(s=(1+1/2)(1-i\varrho(\tau))+i\sigma(\tau))\,\, \nonumber \\
& = &\frac{\eta_D(q)^2}{\eta_D(q^{\frac{1}{2}})\eta_D(q^2)}\,,
\\
\varphi_3(q) & = & \,\, \,\, q^{\frac{1}{24}}\prod_{m=1}^\infty
(1+q^{m+1})\,\, = q^{\frac{1}{24}}\cR(s= 2(1-i\varrho(\tau))+i\sigma(\tau))\,\,\nonumber \\
& = & \frac{\eta_D(q^2)}{\eta_D(q)}\,,
\end{eqnarray}
where
$
\eta_D(q) \equiv q^{1/24}\prod_{n=1}^\infty(1-q^{n})
$
is the Dedekind $\eta$-function. The linear span of $\varphi_1(q), \varphi_2(q)$
and $\varphi_3(q)$ is $SL(2, {\mathbb Z})$-invariant \cite{Kac_book}.
For $b\in {\mathbb R}$ the following relation holds:
\begin{eqnarray}
\prod_{n=\ell}^\infty(1-q^{an+\varepsilon})^{bn} &= &\cR(s=(a\ell+\varepsilon)(1-i\varrho(\tau))+1-a)^{b\ell}
\nonumber \\
&\times & \!\!\prod_{n=\ell+1}^\infty \cR(s= (an+\varepsilon)(1-i\varrho(\tau))+ 1-a)^b.
\end{eqnarray}

\section{Hilbert schemes} \label{Sub-leading}
\label{Hilbert}

{\bf Relevant preliminaries and terminologies.}
A complex $n\times n$ matrix $A = \{a_{ij}\}_{i, j = 1}^n$ of rank $n$
is called a {\it generalized Cartan matrix} if it satisfies the following conditions:
$a_{ii} = 2$ for $i = 1, \cdots, n$;
$a_{ij}$ non-positive integers for $i \neq j$;
$a_{ij} = 0$ implies $a_{ji} = 0$.

A realization of $A$ is a triple $\{{\mathfrak h}, \Pi, \Pi^\vee
\}$, where $\mathfrak h$ is a complex vector space, $\Pi =
\{\alpha_1, \cdots, \alpha_n\}\subset {\mathfrak h}^*$ and
$\Pi^\vee = \{\alpha_1^\vee, \cdots, \alpha_n^\vee\}$ are indexed
subsets in ${\mathfrak h}^*$ and $\mathfrak h$, respectively.

We also set $Q := \sum_{j=1}^n {\mathbb Z}\alpha_j$, \, $Q_+ :=
\sum_{j=1}^n {\mathbb Z}_+\alpha_j$; the lattice $Q$ is called the
{\it root lattice}.
Besides introduce the following root space decomposition with respect to $\mathfrak h$:
${\mathfrak g}(A) = \bigoplus_{\alpha\in Q}{\mathfrak g}_\alpha$, where ${\mathfrak
g}_\alpha = \{x\in {\mathfrak g}(A)\mid [h, x] = \alpha(h)x\,\,\,
\forall h\in {\mathfrak h}\}$ is the root space attached to
$\alpha$. In addition ${\mathfrak g}_0 = {\mathfrak h}$, the
number ${\rm mult}\, \alpha := {\rm dim}\, {\mathfrak g}_\alpha$ is
called the {\it multiplicity} of $\alpha$.
An element $\alpha \in Q$ is called a {\it root} if $\alpha \neq 0$ and ${\rm mult}\,
\alpha \neq 0$; a root $\alpha > 0$ (resp. $\alpha < 0$) is called
{\it positive} (resp. {\it negative}). Denote by $\triangle,
\triangle^+, \triangle^-$ the sets of all roots, positive and
negative roots respectively, such that $\triangle = \triangle^+\cup
\triangle^-$ (a disjoint union).

{\bf The generalized partition function.}
Let ${\mathfrak n}^+$ (resp. ${\mathfrak n} ^-$) denote the
subalgebra of ${\mathfrak g}(A)$ generated by $e_1, \cdots, e_n$
(resp. $f_1, \cdots, f_n$). Then we have the triangular
decomposition: ${\mathfrak g}(A) = {\mathfrak n}^-\oplus
{\mathfrak h}\oplus {\mathfrak n}^+$ (direct sum of vector
spaces). ${\mathfrak g}_\alpha\subset{\mathfrak n}^+$ if $\alpha >
0$ and ${\mathfrak g}_\alpha\subset{\mathfrak n}^-$ if $\alpha <
0$. It means that for $\alpha > 0$ (resp. $\alpha < 0$),
${\mathfrak g}_\alpha$ is the linear span of the elements of the
form $[\cdots [[e_{i_1}, e_{i_2}], e_{i_3}]\cdots e_{i_s}]$ (resp.
$[\cdots [[f_{i_1}, f_{i_2}], f_{i_3}]\cdots f_{i_s}]$) such that
$\alpha_{i_1}+ \cdots + \alpha_{i_s} = \alpha$ (resp. $=
-\alpha$). Besides, ${\mathfrak g}_{\alpha_i} = {\mathbb C}e_i$,
${\mathfrak g}_{-\alpha_i} = {\mathbb C}f_i$, ${\mathfrak
g}_{s\alpha_i} = 0$ if $|s| > 1$. The {\it Chevalley involution}
of the Lie algebra ${\mathfrak g}(A)$ is determined by $w(e_i) = -
f_i,\, w(f_i) = -e_i,\, w(h) = -h$ if $h\in {\mathfrak h}$. Let
$\epsilon(w) \equiv {\rm det}_{{\mathfrak h}^*}w =
(-1)^{\ell(w)}$, where $\ell(w)$ is the length of $w$; also
$w({\mathfrak g}_\alpha) = {\mathfrak g}_{-\alpha}$, ${\rm mult}\,
\alpha = {\rm mult}\, (-\alpha)$ and \, $\triangle^- = -
\triangle^+$. Consider the expression \cite{Kac_book}
\begin{equation}
\prod_{\alpha \in \triangle^+}(1-e(-\alpha))^{- {\rm mult}\, \alpha}
= \sum_{\xi \in {\mathfrak h}^*} {K}(\xi) e(\xi)\,, \label{FK}
\end{equation}
defining a function $K$ on ${\mathfrak h}^*$ called the
(generalized) partition function (the symbol $K$ is in honour of Kostant). Note
that $K(\xi) = 0$, unless $\xi \in Q_+$; furthermore, $K(0) =
1$, and $K(\xi)$ for $\xi \in Q_+$ is the number of partitions
of $\xi$ into a sum of positive roots, where each root is counted
with its multiplicity. Another form of formula  (\ref{FK}) is:
$ \sum_{\xi \in Q_+} K(\xi) e(\xi) = \Pi_{\alpha \in \triangle^+}
(1 + e(\alpha) + e(2\alpha) + \cdots)^{{\rm mult}\, \alpha}.$

Define $\triangle_0 = \{\alpha\in\triangle\mid{\overline \alpha}
=0\}$. The subgroup $W$ of $GL({\mathfrak h}^*)$ generated by all
fundamental reflections is called the Weyl group of ${\mathfrak
g}(A)$. The action of $r_i$ on ${\mathfrak h}^*$ induces the dual
fundamental reflections $r_i^\vee$ on $\mathfrak h$ (for the dual
algebra ${\mathfrak g}({}^tA)$). For each $i=1, \cdots, n$ we
define the fundamental reflection $r_i$ of the space ${\mathfrak
h}^*$ by $r_i(\lambda) = \lambda -\langle\lambda,
\alpha_i^\vee\rangle \alpha_i, \, \lambda\in {\mathfrak h}^*$. It
is clear that $r_i$ is a reflection since its fixed point set is
$T_i = \{\lambda\in {\mathfrak h}^*\mid\langle\lambda,
\alpha_i^\vee\rangle = 0\}$, and $r_i(\alpha_i) = -\alpha_i$. Let
$W_0$ be a (finite) subgroup of $W$ generated by reflections the
$r_\alpha$, with $\alpha\in \triangle_0$. We can describe the integrable
highest-weight modules $L(\Lambda)$ with respect to the algebras
${\mathfrak g}(A)$, where $A$ is a finite type matrix, and use the
specialization formula \cite{Kac_book}:
\begin{equation}
\prod_{\alpha\in \triangle^+\backslash \triangle_0}
(1-e(-{\overline \alpha}))^{{\rm mult}\,\alpha} = \sum_{w\in
W\backslash W_0}\epsilon(w)K(w(\rho))e(\overline{w(\rho)} -
\overline{\rho})\,. \label{Spec}
\end{equation}

\subsection{The Hilbert scheme of points on surfaces}
\label{Scheme}

Let us explain briefly a relation between the Heisenberg algebra and its representations,
and the Hilbert scheme of points. To be more specific: recall that the infinite-dimensional
Heisenberg algebra (or, simply, the Heisenberg algebra) plays a fundamental role in the representation theory
of the affine Lie algebras.
An important representation of the Heisenberg algebra is the Fock space representation on the polynomial ring
of infinitely many variables. The degrees of polynomials (with different degree variables) give a direct sum
decomposition of the representation, which is called weight space decomposition.

The Hilbert scheme of points on a complex surface appears in the algebraic geometry. This scheme
of points decomposes into infinitely many connected components according to the number of points. Betti
numbers of the Hilbert scheme have been computed in \cite{Gottsche}. The sum of the Betti numbers of
the Hilbert scheme of $N$-points is equal to the dimension of the subspaces of the Fock space representation
of degree $N$.

Considering the generating function of the Poincar\'{e} polynomials associated with set of points
one can get the character of the Fock space representation of the Heisenberg algebra.
The character of the Fock space representation of the Heisenberg algebra (in general
the integrable highest weight representations of affine Lie algebras) are known to have modular
invariance as has been proved in {\rm \cite{Kac2}}. This occurrence is naturally explained through the
relation to partition functions of conformal field theory on a torus. In this connection
the affine Lie algebra has close relation to the conformal field theory.

Let ${R} = {\mathbb Q}[x_1, x_2, ...]$
be the polynomial ring of infinite many variables $\{x_j\}_{j=1}^{\infty}$. Define $P[j]$ as $j\partial/\partial x_j$
and $P[- j]$ as a multiplication of $x_j$ for each positive $j$. Then the commutation relation holds:
$
[\,P[i],\, P[j]\,] = i\delta_{i+j, 0}\,{\rm Id}_{R},\,
$
$i, j \in {\mathbb Z}/\{0\}.$ We define the infinite dimensional Heisenberg algebra as a Lie algebra
generated by $P[j]$ and $K$ with defining relation
\begin{equation}
[\,P[i],\, P[j]\,] = i\delta_{i+j, 0} K_{R},\,\,\,\,\,
[\,P[i],\, K\,] = 0,\,\,\,\,\, i, j \in {\mathbb Z}/\{0\}.
\end{equation}
The above ${R}$ labels the representation.
If $1\in {R}$ is the constant polynomial, then $P[i]1 = 0,\, i\in {\mathbb Z_+}$ and
\begin{equation}
{R} =  {\rm Span}\{ P[-j_1] \cdots  P[-j_k]\,1\mid
k\in {\mathbb Z}_+\cup \{0\}, \,\,\, j_1, \dots, j_k \in
{\mathbb Z}_+\}\,.
\end{equation}
1 is a highest weight vector.
This is known in physics as the {\it bosonic Fock space}.
The operators $P[j]\, (j< 0)$ ($P[j]\, (j>0)$) are the {\it creation {\rm (}annihilation{\rm )} operators},
while 1 is  the {\it vacuum vector}.

Define the degree operator
${\cO}: {R}\rightarrow {R}$ by \,
$
{\cO}(x_1^{m_1} x_2^{m_2}\cdot\cdot\cdot \,)\stackrel{def}{=}
(\sum_i i m_i) x_1^{m_1} x_2^{m_2}\ldots
$
The representation ${R}$ has $\cO$ eigenspace decomposition; the eigenspace with eigenvalue $N$ has a basis
$
x_1^{m_1} x_2^{m_2}\cdot\cdot\cdot
(\sum_i i m_i) = N.
$
Recall that partitions of $N$ are defined by a non-increasing sequence of nonnegative integers
$\nu_1\geq \nu_2 \geq \ldots$ such that $\sum_\ell \nu_\ell = N$. One can represent
$\nu$ as $(1^{m_1}, 2^{m_2}, \cdot\cdot\cdot)$ (where 1 appears $m_1$-times, 2 appears $m_2$-times, $\ldots$ in the sequence).
Therefore, elements of the basis corresponds bijectively to a partition $\nu$. The generating function of eigenspace
dimensions, or the {\it character}
in the terminology of the representation theory, is well--known to have the form
\begin{equation}
{\rm Tr}_{R}\, q^{\cO}\, \stackrel{def}{=}
\sum_{N\in {\mathbb Z}_+\cup \{0\}}
q^N {\rm dim}\,\{ r\in {R}\, \mid \, {\cO}r = Nr\,\}
= \prod_{n=1}^\infty (1 - q^n)^{-1} = \cR(s= 1-i\varrho(\tau))^{-1}\,.
\label{character}
\end{equation}

{\bf Application: the Hirota bilinear equation.}
We proceed a basic construction for the Hirota bilinear equation by using of an elements of appropriate
representation theory (detailed description of such a construction the reader can find in
monograph \cite{Kac_book}).
Let $P(x_1,x_2,\ldots)$ be a polynomial depending on a finite number of the $x_j$.
Let $f(x)$ and $g(x)$ be two $C^\infty$-functions.

\begin{definition}
Denote by $P(\mD_1, \mD_2, \ldots)f\cdot g$ the following expression
\begin{equation}
P\left(\frac{\partial}{\partial\xi_1},\frac{\partial}{\partial\xi_2},\ldots\right)f(x_1+\xi_1, x_2+\xi_2, \ldots)
g(x_1-\xi_1, x_2-\xi_2, \ldots)\vert_{\xi=0}.
\end{equation}
The equation $P(\mD)f\cdot g =0$ is called a Hirota bilinear equation.
\end{definition}

If $P=x_1^n$, then using Leibniz's formula we get
\begin{equation}
\mD_1^nf\cdot g = \sum_{k=0}^n (-1)^kC_k^n \frac{\partial^{k}f}{\partial x_1^k}\frac{\partial^{n-k}g}{\partial x_1^{n-k}}.
\end{equation}
$Pf\cdot f\equiv 0$ if and only if $P(x)=-P(-x)$. In this case we have deal with a trivial Hirota bilinear equation.

Denote by $Hir := \oplus_k\cH_k$ the space of Hirota polynomials; $\cH_k\subset {\mathbb C}[y_j;j\in E_+]$ is the space
of principal degree $k$ (the principal degree is defined by $deg\, y_j=j$) and $E_+$ is the sequence of positive exponents
of the Lie algebra $g(A)$ associated to $A$. The straightforward calculations give \cite{Kac_book};
\begin{equation}
dim_q\, Hir := \sum_{n=0}^\infty q^n(dim\,\cH_n)
=  \prod_{n=1}^\infty (1-q^{2n-1})^{-1} - \prod_{n=1}^\infty(1-q^{4n-2})^{-1}.
\end{equation}
Using the Ruelle type spectral functions (see Sect. \ref{Ruelle}) we get
\begin{equation}
 dim_q\, Hir = \cR(s=-i\varrho(\tau))^{-1} - \cR(s=-2i\varrho(\tau)-1)^{-1}
\end{equation}

{\bf The Heisenberg algebra.}
Let us define now the Heisenberg algebra associated with a finite dimensional $\mathbb Q$-vector space $V$ with
non-degenerate symmetric bilinear form $\langle\, ,\, \rangle$.
Let $W = (V\otimes t\,{\mathbb Q}[t])\oplus (V\otimes t^{-1}\,{\mathbb Q}[t^{-1}])$,
then define a skew-symmetric bilinear form on $W$ by
$\langle r\otimes t^i,\,s\otimes t^j\rangle = i \delta_{i+j, 0} \langle r, s\rangle$.

The Heisenberg algebra associated with $V$ can be defined as follows: we take the quotient of the free algebra $L(W)$ divided
by the ideal $\mathcal I$ generated by $[r,\, s] - \langle r,\,s\rangle 1\,\, (r, s\in W)$. It is clear that when $V = {\mathbb Q}$
we have the above Heisenberg algebra. For an orthogonal basis
$\{ r_j\}_{j=1}^n$ the Heisenberg algebra associated with $V$ is isomorphic to the tensor product of $n$-copies
of the above Heisenberg algebra.

Let us consider next the super-version of the Heisenberg algebra, the super-Heisenberg algebra.
The initial data are a vector space
$V$ with a decomposition $V = V_{\rm even}\oplus V_{\rm odd}$
and a non-degenerate bilinear form.
As above we can define $W$, the bilinear form on $W$, and $L(W)/{\mathcal I}$,
where now we replace the Lie bracket $[\, ,\,]$ by the super-Lie bracket. By generalizing the
representation on the space of polynomials of infinite many variables one can get a
representation of the super-Heisenberg algebra on the symmetric algebra ${R}= S^*(V \otimes t\, {\mathbb Q} [t])$
of the positive degree part $V\otimes t\, {\mathbb Q}\,[t]$. As above we can define the degree operator $\cO$.
The following character formula holds:
\begin{equation}
{\rm Tr}_{R}\, q^{\cO} = \prod_{n=1}^\infty
\frac{(1 + q^n)^{{\rm dim}\, V_{\rm odd}}}
{(1 - q^n)^{{\rm dim}\, V_{\rm even}}}
=
\frac{\cR(s = 1-i\varrho(\tau)+ i\sigma(\tau))^{{\rm dim}\, V_{\rm odd}}}
{\cR(s = 1-i\varrho(\tau))^{{\rm dim}\, V_{\rm even}}}\,.
\label{trace-usual}
\end{equation}

\begin{remark}
 In the case when $V$ has one-dimensional odd degree part only
(the bilinear form is $\langle r, r\rangle =1$ for a nonzero vector $r\in V$).
We can modify the definition of the corresponding super-Heisenberg algebra by changing the bilinear
form on $W$ as $\langle r\otimes t^i, r\otimes t^j\rangle = \delta_{i+j, 0}$. The resulting algebra is called
infinite dimensional Clifford algebra. The above representation $R$ can be modified as follows
and it is the fermionic Fock space in physics. The representation of the even degree part was realized
as the space of polynomials of infinity many variables; the Clifford algebra is realized on the exterior algebra
${R} = \wedge^*(\bigoplus_j{\mathbb Q} dx_j)$ of a vector space with a basis of infinity many vectors.
For $j>0$ we define $r\otimes t^{- j}$ as an exterior product of $dx_j$, $r\otimes t^j$ as an interior
product of $\partial/\partial x_j$.
\end{remark}

\subsection{The Verma modules over Virasoro algebras}
\label{Virasoro}

We briefly note  some elements of the representation theory of
Virasoro algebras which are, in fact, very similar to those for
Kac-Moody algebras.
A remarkable link between the theory of highest-weight modules
over the Virasoro algebra, conformal field theory and statistical
mechanics was discovered in \cite{Belavin1,Belavin2,Dowker}.

{\bf Imbedding and couplings in ${\mathfrak g}{\mathfrak l}_{\widehat{\cJ}}({\mathbb K })$ of Kac-Moody and Virasoro algebras.}
We start with very well known Lie algebra ${\mathfrak g}{\mathfrak l}(n, {\mathbb K })$.
The symbol $\mathbb K $ denotes the field of real numbers $\mathbb R$ or the field $\mathbb C$ of complex numbers. In particular,
${\mathfrak g}{\mathfrak l}(n, {\mathbb C})$ is the Lie algebra of all complex $n\times n$ matrices with the operation
$A, B \mapsto [A, B]= AB - BA$.
\begin{remark}
Results for ${\mathfrak g}{\mathfrak l}(n, {\mathbb K })$
survive the passage to the limit $n\rightarrow \infty$, if one assumes that
${\mathfrak g}{\mathfrak l}(\infty, {\mathbb K })$ is the Lie algebra of infinite finitary matrices, it means
$\bigcup_n {\mathfrak g}{\mathfrak l}(n, {\mathbb K })$.
\end{remark}
Now we deal with the Lie algebra
${\mathfrak g}{\mathfrak l}_{\widehat{\cJ}} (\mathbb K )$ of generalized Jacobian matrices.
Note that the bilateral matrix $\Vert a_{ij}\Vert_{i,j\in {\mathbb Z}}$ is called a generalized Jacobian matrix if it has
a finite number of nonzero diagonals (that is, if there exists a positive $N$ such that $a_{ij}=0$ for $\vert j-i\vert> N$).
It is clear that the set of generalized Jacobian matrices constitutes a Lie algebra, with respect to the usual commutation rule.

The algebra ${\mathfrak g}{\mathfrak l}_{\widehat{\cJ}} (\mathbb K )$ can be considered as a nontrivial one-dimensional central
extension of the Lie algebra ${\mathfrak g}{\mathfrak l}_{\cJ}({\mathbb K })$ (for details, see \cite{Fuks}).
It is obvious that ${\mathfrak g}{\mathfrak l}_{\cJ}({\mathbb K })\supset {\mathfrak g}{\mathfrak l}_{\cJ}(\infty, {\mathbb K })$.
It should be pointed out that the importance of the Lie algebra ${\mathfrak g}{\mathfrak l}_{\widehat{\cJ}} (\mathbb K )$
follows from the facts:

-- Many of the classical constructions of the theory of representations
of the Lie algebra ${\mathfrak g}{\mathfrak l}_{\cJ}({\mathbb K })$ can be also applied to the algebra
${\mathfrak g}{\mathfrak l}_{\widehat{\cJ}}({\mathbb K })$. This creates a sizable supply of
${\mathfrak g}{\mathfrak l}_{\widehat{\cJ}}({\mathbb K })$-modules.

-- Important infinite-dimensional Lie algebras can be embedded in ${\mathfrak g}{\mathfrak l}_{\widehat{\cJ}}({\mathbb K })$.
Thus, the already mentioned representations of ${\mathfrak g}{\mathfrak l}_{\widehat{\cJ}}({\mathbb K })$
become representations of these algebras.

-- The subalgebra of ${\mathfrak g}{\mathfrak l}_{\cJ} ({\mathbb K })$ composed of $n$-periodic matrices,
$\Vert a_{ij}\Vert$ with $a_{i+n, j+n}=a_{ij}$, is isomorphic to the algebra of currents
\cite{Fuks}.

Recall that the space of smooth maps $X \rightarrow {\mathfrak g}$,
where $X$ is a smooth manifold and ${\mathfrak g}$ is a finite-dimensional Lie algebra, with the
$\cC^\infty$-topology and the commutator $[f, g](x) = [f(x), g(x)]$, is a (topological) {\it current Lie algebra}
and is denoted by ${\mathfrak g}^X$. Together with the algebra ${\mathfrak g}^{S^1}$ \,($X=S^1$) one can consider its subalgebra
$({\mathfrak g}^{S^1})^{\rm pol}$, consisting of maps described by trigonometric polynomials.
For any commutative associative algebra $A$, the tensor product ${\mathfrak g}\otimes A$ is a Lie algebra with respect
to the commutators $[g_1\otimes a_1, g_2\otimes a_2] = [g_1, g_2]\otimes a_1 a_2$; also
$({\mathfrak g}^{S^1})^{\rm pol} = {\mathfrak g}\otimes{\mathbb C}[t, t^{-1}]$.

A non-trivial central extension of
${\mathfrak g}^X$ -- a Kac-Moody algebra -- is embedded in ${\mathfrak g}{\mathfrak l}_{\widehat{\cJ}} ({\mathbb K })$. The Lie algebra
${L}^{\rm pol} = {\mathbb C}({\rm Vect}\,S^1)^{\rm pol}$ of complex polynomial vector fields on the circle can be embedded in
${\mathfrak g}{\mathfrak l}_{\cJ}({\mathbb K }= {\mathbb C})$. Recall that
${L}^{\rm pol}$ has a basis ${\bf e}_i$ and commutators of the form
\begin{equation}
[{\bf e}_i, {\bf e}_j]  =  (i-j) {\bf e}_{i+j}\,\,\,\,\,
(j\in {\mathbb Z}),
\,\,\,\,\,
{\bf e}_j  =   -z^{j+1}d/dz \,\,\,\,\,
{\rm on}\,\,\,\,\, {\mathbb C}\setminus \{0\}\,.
\label{basis}
\end{equation}
(The cohomologies of the algebra ${L}^{\rm pol}$ are known; in particular, $H^2({L}^{\rm pol})= {\mathbb C}$.)
The Virasoro algebra is a Lie algebra over $\mathbb C$. Because of
Eq.~(\ref{basis}), the Lie Virasoro algebra is a (universal) central extension of the Lie algebra of holomorphic
vector fields on the punctured complex plane having finite Laurent series. For this reason the Virasoro
algebra plays a key role in conformal field theory.

{\bf Representations of the Virasoro algebra.}
Here we briefly note  some elements of the representation theory of
Virasoro algebras which are, in fact, very similar to those for Kac-Moody algebras.
Consider the highest representation of
the Virasoro algebra. Let $M(c, h)\, (c, h \in {\mathbb C})$ be
the Verma module over the Virasoro algebra.
The {\it conformal central charge} $c$ acts on $M(c,
h)$ as $cId$. As $[{\bf e}_0, {\bf e}_{-j}] = n {\bf e}_{-j}$,
${\bf e}_0$ is diagonalizable on $M(c, h)$, with spectrum $h+
{\mathbb Z}_{+}$ and eigenspace decomposition given by: $ M(c, h)
=\bigoplus_{j\in {\mathbb Z}_{+}} M(c, h)_{h+j}, $ where $M(c,
h)_{h+j}$ is spanned by elements of the basis $\{{\bf
e}_{-j_k}\}_{k=1}^n$ of $M(c, h)$. The number $ Z_j = {\rm dim}\,
M(c, h)_{h+j}, $ is the {\it classical partition function}. This
means that the Konstant partition function for the Virasoro
algebra is the classical partition function. On the other hand,
the partition functions can be rewritten in the form (Cf. Eq. (\ref{character}))
\begin{eqnarray}
{\rm Tr}_{M(c, h)}\, q^{{\bf e}_0} &:=& \sum_{\lambda}{\rm dim}\,
M(c, h)_{\lambda}\,q^{\lambda} = q^h\prod_{j=1}^\infty (1-q^j)^{-1}
\nonumber \\
&=& q^h {\cR}(s= 1-i\varrho(\tau))^{-1}.
\label{ch}
\end{eqnarray}
The series ${\rm Tr}_{M(c, h)}\,q^{{\bf e}_0}$ is called the formal character of the Virasoro (Vir)-module ${M(c, h)}$.
\begin{note}
A $\mathfrak g$-module ${V} \in {C}$, where $C$ is a category, if: there is an expansion ${V}=
\bigoplus_{\lambda \in {\mathfrak h}^*}{V}_\lambda$\, ($\mathfrak h$ is a Cartan subalgebra of $\mathfrak g$)
and $e_\alpha^{(i)}{V}_\lambda \subset {V}_{\lambda+\alpha}$, where $e_\alpha^{(i)}$ are root vectors correspond
to root $\alpha$; ${\rm dim}\,{V}_\lambda < \infty$ for all $\lambda$; $D(\lambda) := \{ \lambda \in
{\mathfrak h}^*\mid {V}_\lambda \neq 0\}\subset \bigcup_{i=1}^sD(\lambda_{i})$ for some
$\lambda_1, \ldots, \lambda_s \in {\mathfrak h}^*$.
\end{note}
As a result the analytic structure of the formal character of the $Vir$-module ${M(c, h)}$ is
determined by the pole structure of the Ruelle-type spectral function.

\section{Invariants from algebraic geometry}
\label{Invariants}

\subsection{The analytic torsion \texorpdfstring{$\tau^{\bf an}$}{} and
\texorpdfstring{$\eta$}{eta}-invariant via cyclic homology}
\label{cyclic}

{\bf Cyclic (co)homology of regular algebras of smooth manifolds.}
The definition of the $\eta$-invariant of $X$ and the R-torsion can be reformulated
in terms of the cyclic homology and homology of the Lie algebra $\cA = C^{\infty}(X)$.
We begin with this reformulation.

Let $C_\ast(\cA)$ denote the cyclic object associated with the standard complex of and $\cA$
over $k$. The standard double complex $CC_{\ast\ast}(\cA)$ associated with this cyclic object
and hense also with $\cA$ is the first quadrant double complex.
By definition, the cyclic homology $HC_\ast(\cA)$ of $\cA$ is the homology
$H_ \ast(CC_\ast(\cA))$ of the standard total complex of the standard double complex of $\cA$.
The standard double complex $CC_{ \ast \ast}(\cA)$, its associated total complex
$CC_\ast(\cA)$, and the cyclic homology $HC_\ast(\cA)$, are all functors of $\cA$
on the category of algebras over $k$, since the standard cyclic object $C_\ast(\cA)$
is functorial in $\cA$ from the category of algebras over $k$ to the category of
cyclic $k$-modules $\Lambda(k)$.

There is exact cohomological Connes sequence
\begin{equation}
\cdots \longrightarrow HH^n(\cA) \stackrel{B}{\longrightarrow}
HC^{n-1}(\cA)\stackrel{S}{\longrightarrow}HC^{n+1}(\cA)\stackrel{I}{\longrightarrow}
HH^{n+1}(\cA)\longrightarrow \cdots
\end{equation}
Since of duality properties for cyclic cohomology and homology we have
\begin{equation}
\cdots \longrightarrow HH_n(\cA) \stackrel{I}{\longrightarrow}
HC_n(\cA)\stackrel{S}{\longrightarrow}HC_{n-2}(\cA)\stackrel{B}{\longrightarrow}
HH_{n-1}(\cA)\longrightarrow \cdots
\end{equation}
Let $M_r(\cA)$ be an algebra of matricies of dimension $r$ with coefficients from $\cA$. Then for
$\alpha = \{\alpha_{ij}\}_{i,j=1}^r\in M_r(\cA)$ assume ${\rm Tr}(\alpha)=\sum_{i=1}^r\alpha_{ii}$ and
define
\begin{equation}
 {\rm Tr}: M_r(\cA)^{\otimes(n+1)}\longrightarrow \cA^{\otimes(n+1)}\,\,\,\,\, {\rm by}
\,\,\,\,\,{\rm formula}\,\,\,\,\,
 {\rm Tr}(\alpha\otimes \beta\otimes \ldots \otimes\gamma) =
\sum \alpha_{i_0i_1}\otimes \beta_{i_1i_2}\otimes \ldots \otimes\gamma_{i_ni_0},
\nonumber
\end{equation}
where the sum is taken over all sets of indixes $(i_0,i_1, \ldots , i_n)$.
Due to natural isomorphism $M_r(\cA)\cong M_r(k)\otimes \cA$, any element from $M_r(\cA)$
represents in the form of sum of elements $xa$, $x\in M_r(k)$, $a\in \cA$, while an element
from $M_r(\cA)^{\otimes(n+1)}$ represents as sum of elements
$x_0a_0\otimes \ldots \otimes x_na_n$, where $x_i\in M_r(k)$, $a_i\in \cA$.
The trace mapping has the form
\begin{eqnarray}
 {\rm Tr}(x_0a_0\otimes \ldots \otimes x_na_n) & = &
\sum (x_0 a_0)_{i_0i_1}\otimes \ldots \otimes (x_n a_n)_{i_ni_0}
\nonumber \\
& = &
\sum (x_0)_{i_0i_1}\ldots (x_n)_{i_ni_0} a_0\otimes \ldots \otimes a_n
\nonumber \\
& = &
 {\rm Tr}(x_0\ldots x_n)a_0\otimes \ldots \otimes a_n\,.
\end{eqnarray}

It can be shown that trace mapping is the morphism of chain complexes
$ {\rm Tr}: C_\ast(M_r(\cA))\rightarrow C_\ast(\cA)$; it induces isomorphism
(Morita equivalence for matricies) \cite{Troitsky}
\begin{equation}
\,\, {{\rm Tr}_\ast}: HH_\ast(M_r(\cA))\stackrel{\cong}{\longrightarrow}
HH_\ast(\cA),\,\,\,\,\,\,
\,\, HH^\ast(M_r(\cA))\stackrel{\cong}{\longrightarrow}
HH^\ast(\cA),
\end{equation}
where by definition $HH^n(\cA)= H^n(C^\ast(\cA))= H^\ast({\rm hom}(C_\ast(\cA), k))$.
Define pairing of cochains and chains
\begin{equation}
\langle \cdot , \cdot \rangle: C^n(\cA)\times C_n(\cA)\longrightarrow k
\end{equation}
Since $\langle b^\ast(f), x\rangle = \langle f, b(x)\rangle$, $f\in C^n(\cA)$, $x\in C_{n+1}(\cA)$
the mapping $\langle \cdot , \cdot \rangle$ induces the following pairing:
\begin{equation}
\langle \cdot , \cdot \rangle: HH^n(\cA)\times HH_n(\cA)\longrightarrow k\,\,\,\,\,
({\rm Kronecker \,\,\,\, product})
\end{equation}
In the case of chains we have $\langle b^\ast(f), x \rangle = \langle f, b(x)\rangle$
for any $f\in C^n(\cA)$ and $x\in C_{n-1}(\cA)$, and therefore
\begin{equation}
\langle \cdot , \cdot \rangle: HC^n(\cA)\times HC_n(\cA)\longrightarrow k\,.
\end{equation}

Let $\cA$ be a regular $k$-algebra over a field $k$
of characteristic zero. Then the cyclic homology is given by
\begin{equation}
HC_p(\cA) = \Omega^p_{\cA/k}/d\Omega^{p-1}_{\cA/k}\oplus H^{p-2}_{DR}(\cA)
\oplus H^{p-4}_{DR}(\cA)\oplus \cdots
\label{HC1}
\end{equation}
Let $\cA$ be the $\mathbb C$-algebra of smooth functions on a smooth manifold.
Then the cyclic homology is
\begin{equation}
HC_p(\cA) = \cA^p(X)/d\cA^{p-1}(X)\oplus H^{p-2}_{DR}(X)
\oplus H^{p-4}_{DR}(X)\oplus \cdots
\label{HC2}
\end{equation}
In both cases, (\ref{HC1}) and (\ref{HC2}), the projection of $HC_p(\cA)$ onto the first
term is induced by a morphism $\mu: (C_\ast(\cA), b, B)\rightarrow \Omega^\ast(\cA/k, 0, d)$.

{\bf Homomorphism $I$}: $HH_p(\cA)\rightarrow HC_p(\cA)$ is the projection of $HH_p(\cA)$, the $p$-forms,
onto the first factor of $HC_p(\cA)$.

{\bf Homomorphism $S$}: $HC_p(\cA)\rightarrow HC_{p-2}(\cA)$ is injection of the first factor of
$HC_p(\cA)$ into the second factor $H^{p-2}_{DR}$ and the other factors map isomorphically on the
corresponding factor of $HC_{p-2}(\cA)$.

{\bf Homomorphism $B$}: $HC_{p-2}(\cA)\rightarrow HH_{p-1}(\cA)$ is zero on all factors except the first
one where it is $d: \Omega^{p-2}/d\Omega^{p-3}\rightarrow \Omega^{p-1}$.

These results follows from the fact that we can calculate cyclic homology,
Hochschild homology, and the Connes exact couple with the mixed complex
$(\Omega, 0, d)$ and it is an easy generality on mixed complexes with the first
differential zero.

{\bf Cyclic homology and the Connes exact couple.}
From the 2-fold periodicity of the double complex $CC_{ \ast \ast}(\cA)$, we have a morphism
$CC_{ \ast \ast}(\cA)\rightarrow CC_\ast(\cA)$ of bidegree (2, 0), giving a
morphism $\sigma : CC_\ast(\cA)\rightarrow CC_\ast(\cA)$ of degree 2 and a short
exact sequence of complexes
\begin{equation}
0\longrightarrow {\rm ker}(\sigma)\longrightarrow CC_\ast(\cA)
\stackrel{\sigma}{\longrightarrow} CC_\ast(\cA)\longrightarrow 0\,.
\label{SS}
\end{equation}
Then we have $H_ \ast({\rm ker}(\sigma)) = HH_ \ast(\cA)$. The homology exact triangle
of the short exact sequence of complexes (\ref{SS}) is the Connes exact triangle,
where $S = H_\ast(\sigma)$ such that ${\rm deg}(S) = -2$, ${\rm deg}(B) = +1$,
and ${\rm deg}(I) = 0$. This defines a functor from the category of algebras over $k$
to the category of positively $\mathbb Z$-graded exact couples ${\rm Ex}C(-2, +1, 0)$
over the category of $k$-modules $(k)$.

\begin{itemize}
\item{}
$HH_\ast(\cA)$ is canonically isomorphic to $\Omega^*(X)$ (Hochschild, Kostant, and Rosenberg
theorem); under this isomorphism, the
exterior derivative $d : \Omega^n(X) \rightarrow \Omega^{n+1}(X)$ corresponds to the operator
$\beta = B\circ I: HH_n(\cA)\rightarrow HH_{n+1}(\cA)$.

\item{}
$HC_n(\cA) \cong \Omega^n(M)/d\Omega^{n-1}(M)\oplus H^{n-2}(M)\oplus H^{n-4}(M)
\oplus \cdots$
By transport of structure, both $HH_\ast (\cA)$ and $HC_\ast (\cA)$ become (graded)
pre-Hilbert spaces.
\end{itemize}

\begin{Resume}
Note that $\eta$ can be viewed (apart of zero modes) as the regularized signature of
$
\ast \beta: HC_{2k-1}(\cA)  \longrightarrow  HC_{2k-1}(\cA),
$
in addition ${ Tr}: HC_{2k-1}(M_r(\cA))  \stackrel{\cong}{\longrightarrow} HC_{2k-1}(\cA)\,.
$

We should note that a similar interpretation is also exists for the analytic torsion $\tau^{\bf an}$.
Indeed, regarding
$
{\rm log}\,{\rm det}_\zeta ((\ast\beta)\beta:
HC_n(\cA) \rightarrow HC_n(\cA)),
$
(${\rm det}_\zeta$ denotes the $\zeta$-determinant of
the corresponding positive elliptic operator with zero modes removed),
as a regularized dimension of $HC_q(\cA)$, the above
formula identifies ${\rm log}\,\tau$ as the regularized Euler characteristic of $HC_\ast(\cA)$.
The analytic torsion can be interpreted as a regularized pairing  \cite{Moscovici}:
$\langle e(X, g), [X, g]\rangle_\rho$,  ($e(X, g)$ denotes the Euler class),
depending on the representation $\rho$, the regulariozation being given by the geodesic spectrum.

We recall that $\eta_\rho$ and  ${\rm log}\,\tau^{\bf an}_\rho$ is also can be defined
by a regularization procedure involving the ``dual'' data, namely the spectrum of elliptic
operators. Let us explain this statement explicitly.

{\bf $\eta(s, {\fD})$-invariant.}
For a self-adjoint elliptic differential operators acting on a complex manifold $X$
the eta-invariant has been introduced in \cite{APS}, closely connected to the index theorem
for manifolds with boundary. Note that it is a spectral invariant, it measures the asymmetry
of the spectrum {\it Spec}($\fD$) of an opprapriate operator $\fD$.
On behalf of definition such an operator for $Re(s)>>0$ is
\begin{eqnarray}
\eta(s, {\fD}) & = & \sum_{\lambda \in Spec(\fD)/\{0\}}sgn\lambda\vert\lambda\vert^{-s}
= {\rm Tr}   ({\fD}({\fD}^2)^{-\frac{s+1}{2}})
\nonumber \\
& = & \frac{1}{\Gamma(\frac{s+1}{2})}\int_0^\infty  t^{\frac{s-1}{2}}{\rm Tr} ({\fD}\exp(-t{\fD}^2))dt
\end{eqnarray}
$\eta(s, {\fD})$ admits a meromorphic extension to the whole $s$-plane with at most simple poles
at $s= (dim\, X-k)/(ord\, \fD)$, where $k=0, 1, 2, \ldots$. For these poles residues are locally
computable. The point $s=0$ is not a pole, therefore this maces it possible to define the eta-invariant
of $\fD$ by setting $\eta({\fD}) = \eta(0, {\fD})$.
Note that one can attach an eta-invariant to any Dirac type operator on a compact Riemannian manifold
of odd dimensional. On even dimensional manifold Dirac operators have symmetric spectrum and
trivial eta-invariants. We note also the even part of such an operator, the tangential signature operator
$\mathcal B$. The eta-invariant  of $X$ takes the form $\eta_{X} = \eta({\mathcal B})$.

{\bf Determinant of a finite collection of positive defined operators.}
Let $\fD$ be a  unbounded self-adjoint operator on a separable Hilbert space and let $\fD$ be an invertible
operator with $p$-summable inverse :
\begin{equation}
{\rm Tr}\, {\fD}^{-p}<\infty, \,\,\,\, {\rm for}\,\,\,\, {\rm some}\,\,\,  p\in [1, \infty).
\label{operator}
\end{equation}
Define the associate theta-function $\vartheta_{\fD}(t) = {\rm Tr}e^{-t{\fD}}, t>0$, and suppose that the following
expansion holds:
\begin{equation}
\vartheta_{\fD}(t)\vert_{t\rightarrow 0^+}\sim \sum_{n=0}^\infty c_{\alpha_n}t^{\alpha_n},
\label{expansion}
\end{equation}
where $\alpha_j\in{\mathbb R}, -p = \alpha_0<\alpha_1<\cdots < \alpha_j< \cdots, \alpha_j\rightarrow \infty$.
\end{Resume}
 $\vartheta$-regularized determinant of $(\id +\lambda {\fD}^{-1})$ we define as the exponential
 of the meromorphic function ${\rm log}{\rm det}_\vartheta(\id +\lambda {\fD}^{-1})$ which can be uniquely characterized
 by the following requirements:
\begin{eqnarray}
&& \frac{d}{d\lambda}{\rm log}{\rm det}_\vartheta(\id +\lambda {\fD}^{-1}) =  {\rm Tr}_\vartheta ({\fD}+\lambda)^{-1},
\nonumber \\
&& {\rm log}{\rm det}_\vartheta(\id +\lambda {\fD}^{-1})\vert_{\lambda = 0}  =  0.
\nonumber
\end{eqnarray}
Then we need zeta-regularizad determinant ${\rm det}_\zeta$ \cite{RS,RS2} relate to  ${\rm det}_\vartheta$.
For this reason recall that by definition ${\rm log}{\rm det}_\zeta {\fD} = -\frac{d}{ds}\zeta_{\fD}(s)\vert_{s\rightarrow 0}$,
where the zeta-function of ${\fD}$ is $\zeta_{\fD}(s) = {\rm Tr}\,{\fD}^{-s}$ for $Re s >p$.
Also ${\rm log}{\rm det}_\zeta ({\fD}+\lambda) = -\frac{d}{ds}\zeta_{\fD}(s,\lambda)\vert_{s\rightarrow 0}$.
The zeta torsion function we define as  $\tau_\zeta(\lambda) = \tau^{\bf an}_\rho(X;\lambda)$.

 Let us consider  a finite collection of operators ${\fD}_1, \ldots, {\fD}_{2m},\, m\in{\mathbb Z}_+$ and suppose that each
 operator satisfying Eqs (\ref{operator}) and (\ref{expansion}). Let $X$ be a closed, oriented Riemannian manifold
 and $\rho$ be an orthogonal representation: $\pi_1X\rightarrow O(n)$ (see for detail \cite{Fried}).

 One can obtain a complex of an appropriate forms and then the Hodge theory with the twisted cohomology $H^j(X; \rho)$.
 When these cohomology groups vanish for all $j$, an appropriate (orthogonal) representation is acyclic. In this case
 we may define the analytic torsion $\tau^{\bf an}_\rho(X;\lambda)$ by
 \begin{equation}
\tau^{\bf an}_\rho(X;\lambda)^2 =  \frac{\!\!\prod_{j=1,3,5,\ldots}^{2m-1} {\rm det}_\zeta({\fD}_j+\lambda)^j}
{\prod_{j=2,4,6,\ldots}^{2m} {\rm det}_\zeta({\fD}_j+\lambda)^j}.
 \end{equation}

 \subsection{The Riemann-Roch and Atiyah-Singer formulas}
\label{Riemann-Roch}

 Let $\zeta : H\rightarrow K$ be a multiplicative map from one multiplicative theory of
cohomology to another one. (As a simple example we let $\zeta := Ch$.)
Let $\cE$ be a vector bundle, oriented with respect to theory $K$ (but can be non-oriented with
respect to theory $H$).

Note that complex vector bundles at the same time are spinor bundles. Therefore
the characteristic classes of spinor bundles are also characteristic classes of
complex vector bundles and they must be expressed by means of Chern classes.
One can find such expressions if compute classes $c, p_i, \chi$ for the bundle
$E\otimes E \otimes \cdots \otimes E$ over
$(\underbrace{{\mathbb C}P^N\times \cdots \times{\mathbb C}P^N}_n)$.
As a result we have
\begin{eqnarray}
c(E\otimes E\otimes \cdots \otimes E) & = & x_1+ x_2 + \cdots + x_n\,,
\nonumber \\
p_i(E\otimes E\otimes \cdots \otimes E) & = & e_i(x_1^2, x_2^2, \cdots , x_n^2)\,,
\nonumber \\
\chi(E\otimes E\otimes \cdots \otimes E) & = & x_1x_2\cdots x_n\,,
\end{eqnarray}
where $e_i$ are symmetric polynomials ($c,p_i, \chi$ can be expressed in terms of $c_i$).
For the Todd class we have
\begin{eqnarray}
{ Td}(\cE) & = & \exp(c/2)A(p_1(\cE),\, p_2(\cE), \ldots)\,,
\\
A(e_1(x_1^2, x_2^2, \ldots),\, e_2(x_1^2, x_2^2, \ldots), \ldots)
& := & \prod (x_i/2)^{-1}{\rm ch}(x_i/2)\,.
\label{cc}
\end{eqnarray}
The characteristic class $A(p_1(\cE), \, p_2(\cE), \ldots)$ is called the ({\it reducible})
{\it class of Atiyah-Hirzebruch} of real bundle $\cE$ \cite{Karoubi}.

Let ${\mathfrak t}_K$ and ${\mathfrak t}_H$ be a Thom isomorphism and Thom
homomorphism respectively (both possess multiplicative properties). Then for any
$h \in H({X})$ (${X}$ is a base of bundle $\cE$) we have
${\mathfrak t}_K^{-1}\zeta {\mathfrak t}_H(h) = \zeta (h)Td(\cE)$.
Let $E$ be a complex spinor vector bundle with finite CW base ${X}$. Then for any
$h\in K(X)$,
\begin{equation}
Ch({\mathfrak t}_K(E) h) = {\mathfrak t}_H(E(Ch(h)\cdot \exp(c(E)
A(p_1(E)), \, p_2(E), \ldots)\,.
\end{equation}
Let $X$ and $Y$ be manifolds and let their normal bundles $\nu_{\,X}$ and $\nu_{\,Y}$
assume complex spinor structures. Let us consider a continuous map $f: X\rightarrow Y$.
Then for any $h\in K(X)$ we have \cite{FF}
\begin{equation}
f_! ( Ch(h)\cdot \exp (-c(\nu_{\,X}A(\overline{p}_i(X)))
=  Ch(f_!h)\cdot \exp(-c(\nu_{\,Y}A(\overline{p}_i(Y)))\,,
\end{equation}

where $\overline{p}_i(X) = p_i(\nu_{X})$ and $\overline{p}_i(Y) = p_i(\nu_{\,Y})$ are
{\it the Pontriagin normal classes} of manifolds $X$ and $Y$ respectively.
The following formula holds:
\begin{equation}
{\rm Index}\, {\fD} = \langle ( Ch(E)\cdot  Td^{-1}({\mathbb C}
\zeta (X))),\,[X] \rangle\,.
\label{RR}
\end{equation}
Eq. (\ref{RR}) recalls the Riemann-Roch theorem. In (\ref{RR}) $ Td({\mathbb C}\mu(X))$
denotes the Todd class of tangent bundle complexification,
\begin{eqnarray}
 Td^{-1}({\mathbb C}T({X}))  & = &  U(p_1(X), \, p_2(X), \ldots )\,,
\nonumber \\
U(e_1(x_i^2), \, e_2(x_i^2), \ldots) & = &
-\prod x_i^2\left((1-e^{-x_i})(1-e^{x_i})\right)^{-1}\,.
\nonumber
\end{eqnarray}

For any oriented vector bundle $\cE$ define its complexification ${\mathbb C}\cE$,
$\kappa: T(\cE)\rightarrow {\mathbb C}T(\cE)$. This imbedding is induced by natural imbedding
$\cE\rightarrow {\mathbb C}\cE$. For any $h\in H^ \ast(T({\mathbb C\cE}))$ we obtain
\begin{equation}
({\mathfrak t}_H(\cE))^{-1} \kappa^\ast h =
({\mathfrak t}_H({\mathbb C}\cE))^{-1}(h)\cdot \chi(\xX)\,.
\label{chi}
\end{equation}

Using Eq. (\ref{chi}) and the Atiyah-Singer formula we get the final result
(for detail of a computation see \cite{FF}):
\begin{eqnarray}
{\rm Index}\,{\fD} & = &  \langle ( Ch(E)\cdot  {Td}^{-1}({\mathbb C}
\zeta^\ast (X))),\,[X] \rangle
\nonumber \\
& \stackrel{{\rm Eq.}\, (\ref{chi})}{=\!=\!=\!=\!=}&
\langle \chi (\zeta^\ast(X))\cdot {Td}({\mathbb C}\zeta^\ast(\xX))\cdot
{Td}^{-1}({\mathbb C}\zeta^\ast)(X)), \, [X] \rangle
\nonumber \\
& = &
\langle \chi (\zeta^\ast(X)),\, [X] \rangle = \chi(X)\,.
\label{Euler}
\end{eqnarray}
In Eq. (\ref{Euler}) $\chi(X)$ is the Euler characteristic of $X$.
It is known that if $X$ is any compact, connected complex manifold, then
$\chi(X) = {Td}(X)$.

By means of Atiyah-Singer formula one can calculate the index of an elliptic
operator and expess it (i.e. express signature of a manifold) in terms of Pontrjagin
classes.

\subsection{Adiabatic limit and the Dirac index}
\label{Adiabatic}

Eta-invariants of Dirac operators are related to various important invariants from differential topology:
(i) the Chern-Simons invariants, (ii) the Adams e-invariant, and (iii) the Eels-Kuiper invariant
\cite{APS}. Also they related to global anomalies in gauge theories \cite{Witten85}.

{\bf Adiabatic limit.}
Let us consider a determinant construction for a self-adjoint operator
on a finite dimensional Hilbert space.
The classical Cayley transform \cite{Cayley}
for such an operator $\mD$ is the unitary operator $C = (\mD-i)/(\mD + i)$.
For $s\in {\mathbb C}$ we have a family of operators
\begin{equation}
C(s) = \frac{\mD-is}{\mD+is}\,.
\end{equation}
This family is meromorphic, has poles at $s\in i{\rm Spec}^{\prime}({\mathfrak D})$\,
$({\rm Spec}^{\prime}(\mD)\equiv{\rm Spec}({\mathfrak D})-\{0\})$, these poles being simple
and having residue ${\rm Res}_{-i\lambda}C(s) = 2i\lambda P_\lambda$, where $P_\lambda$
is projection onto the $i\lambda$ eigenspace. One has (see for detail \cite{Moscovici89}):
\begin{equation}
{\rm log}\,{\rm det}^\prime C(s) = \!\sum_{\lambda \in {\rm Spec}^\prime (\mD)}
m(\lambda){\rm log} \left(\frac{\lambda -is}{\lambda+ is}\right),
\end{equation}
where $m(\lambda)$ denote the multiplicity.

Let $\mD$ be a Dirac operator, as defined above; the
family of operators $C(s)= (\mD)-is)/(\mD+is)$ is meromorphic with simple poles at
$s\in i{\rm Spec}^\prime (\mD)$. The determinant satisfies the functional identity
$
{\rm det}^\prime(\mD+ is)/(\mD-is)\cdot {\rm det}^\prime (\mD-is)/(\mD+ is) = 1.
$
The following result holds \cite{Moscovici89} (Proposition 2.2):
$
\lim_{\stackrel{x\rightarrow +\infty}{}}
\, {\rm det}^\prime\, C(x) = e^{-i\pi\eta(0, \mD)}.
\label{adiabatic1}
$

Suppose $\varepsilon = x^{-1}$. Then if one replaces the metric $g$ on $X$ by
$g_\varepsilon = g\varepsilon^{-1}$,
the adiabatic limit ({\bf a}--{\rm lim}) of the Cayley
transform of $\mD_\varepsilon$ is $\exp (-i\pi\eta(0, \mD))$,
\begin{equation}
{\bf a}\!-\!\lim_{\stackrel{\varepsilon\rightarrow 0}{}}
\, {\rm det}^\prime\, \frac{\mD - i\varepsilon^{-1}}{\mD +i\varepsilon^{-1}}
= e^{-i\pi\eta(0, \mD)}.
\label{adiabatic2}
\end{equation}

{\bf The Dirac index.}
The eta-invariant  of $X$ plays important role in the index theorem for manifolds with boundary, where
they contribute to the non-local boundary correction terms.
Let ${P}={X}\times{G}$ be a trivial principal bundle over ${X}$
with the gauge group ${G}=SU(n)$ and let $\Omega^1({X};{\mathfrak g})$ be the space
of all connections on ${P}$; this space is an affine space of one-forms on ${X}$ with
values in the Lie algebra ${\mathfrak g}$ of ${G}$.

The holonomies of the parallel transport of flat connections on $P$ give
the identification of ${\mathcal X}_X$ with the space of conjugacy classes of
representations of  $\pi_1(X)$ into $G$, since any principal $G$-bundle $P$ over
a compact oriented three-manifold $X$ is trivial \cite{Nishi}.

We use the notation $CS(\rho) := CS(A_\rho)$
for a representation $\rho$ of $\pi_1(X)$, where $A_\rho$ is a flat connection
corresponding to a representation $\rho$. This gives a topological invariant
for a pair $(X, \rho)$. Since the Chern-Simons invariant is additive with respect to
the sum of representations, we have: $CS(\rho_1\oplus \rho_2) = CS(\rho_1) + CS(\rho_2)$.

Let $X$ be a compact oriented hyperbolic three-manifold, and $\rho$ be an irreducible
representation of $\pi_1(X)$ into $SU(n)$. Denote the corresponding flat vector
bundle by $E_\rho$, and a flat extension of $A_\rho$ over $M$ ($\partial M = X$)
corresponding to $\rho$ by $\underline{A}_\rho$.
The second Chern
character ${ Ch}_2({{\underline E}}_{\rho})\, = CS({\underline A})$
of ${ {\underline  E}}_{\rho}$ can be
expressed in terms of the first and second Chern classes
$
{Ch}_2({ {\underline  E}}_{\rho})= (1/2)c_1({
{\underline E}}_{\rho})^2 - c_2({ {\underline   E}}_{\rho})\,,
$
while the Chern character is
$
{Ch}({ {\underline   E}}_{\rho})  =  {\rm rank}\,
{ {\underline   E}}_{\rho}+
c_1({ {\underline   E}}_{\rho})
+{Ch}_2({ {\underline   E}}_{\rho})
={\rm dim}\,\rho +c_1({ {\underline   E}}_{\rho})+
{Ch}_2({ {\underline   E}}_{\rho})\,.
$

The Atiyah-Patodi-Singer result for manifold with boundary is given by
\cite{APS}:
\begin{equation}
{\rm Index}\, {\mathfrak D}_{\rho} =\int_{M}
{ Ch}
({ {\underline   E}}_{\rho})
{\widehat  A}({M})-\frac{1}{2}(\eta(0,{\mathfrak D}_{\rho})+
h(0,{\mathfrak D}_{\rho}))\,,
\label{Index}
\end{equation}
Here ${\widehat A}({M}) \equiv {\widehat A}(\Omega({M}))$-genus is the usual
polynomial in terms of the Riemannian curvature $\Omega({M})$
of a four-manifold ${M}$ with boundary
$\partial {M} = {X}$. It
is given by ${\widehat  A}({M})= \left({\rm det}\left(\frac{\Omega({M})/4\pi}{{\rm sinh}
\, \Omega({M})/4\pi} \right)\right)^{1/2}$ $= 1- (1/24)p_1({{M}})$, where
$p_1(M) \equiv p_1(\Omega({{M}}))$ is the first Pontrjagin class.
$h(0,{\mathfrak D}_{\rho})$ is the dimension of the space of harmonic
spinors on ${X}$ ($h(0,{\mathfrak D}_{\rho})
={\rm dim}{\rm Ker}\,{\mathfrak D}_{\rho}$ =
multiplicity of the 0-eigenvalue of ${\mathfrak  D}_{\rho}$ acting
on ${X}$ with coefficients in $\rho$).

\subsection{Twisted Dirac operators on locally symmetric spaces}
\label{Twisted}

Let $X := \overline{X}/\Gamma$, $\overline{X}$ is a globally symmetric space of non-compact type
and $\Gamma$ is a discrete,  torsion-free, co-compact subgroup of orientation-preserving isometries.
Thus $X$ inherits a locally symmetric
Riemannian metric $g$ of non-positive sectional curvature. In addition the connected
components of the periodic set of the geodesic flow $\Phi$, acting on the unit tangent
bundle $TX$, are parametrized by the non-trivial conjugacy classes $[\gamma]$ in
$\Gamma = \pi_1(X)$. Therefore each connected component $X_\gamma$ is itself a
closed locally symmetric manifold of non-positive sectional curvature.

\begin{proposition}
Let $\varphi: \Gamma\rightarrow U({F})$ be a unitary representation of $\Gamma$ on $F$.
The Hermitian vector bundle ${E}= {X}\times_{\Gamma}{F}$ over ${X}$
inherits a flat connection from the trivial connection on
$\overline{X}\times {F}$. For any vector bundle $E$ over $X$ let $\overline{E}$ denote
the pull-back to $\overline{X}$.
If ${\mD}: C^{\infty}({X}, V)\rightarrow C^{\infty}({X},V)$ is a differential
operator acting on the sections of the vector bundle $V$, then
${\mathfrak D}$ extends canonically to a differential operator
${\mathfrak D}_{\varphi}: C^{\infty}({X},V\otimes F)\rightarrow
C^{\infty}({X},V\otimes{F})$, uniquely
characterized by the property that ${\mathfrak D}_{\varphi}$ is
locally isomorphic to ${\mathfrak D}\otimes \cdots \otimes {\mathfrak
D}$\,\,\, {\rm (}${\rm dim}\,{F}$ times{\rm )}.

In connection with a real compact hyperbolic manifold $X$ consider a locally homogeneous
Dirac bundle $E$ over $X$ and the corresponding Dirac operator
$\mD: C^\infty(X, E)\rightarrow C^\infty(X, E)$.
As before, assume that $X = \partial M$, that $E$ extends to a Clifford bundle on
$M$ and that $\varphi: \pi_1(X)\rightarrow U(F)$ extends to a representation of
$\pi_1(M)$. Let $\underline{A}_\varphi$ be an extension of a flat connection $A_\varphi$
corresponding to $\varphi$.
\end{proposition}

The Chern-Simons invariant admits a representation in terms of function
$Z(s, \mD_\varphi)$, which is a meromorphic function on $\mathbb C$. $Z(s, \mD_\varphi)$
For ${\rm Re} (s^2)\gg 0$ $Z(s, \mD_\varphi)$ has a meromorphic continuation given by the identity
\cite{Moscovici89,Moscovici91}
\begin{equation}
Z(s,{\mathfrak D_\varphi})= {\rm det}^{\prime}
\left(\frac{{\mathfrak D}_\varphi- is}{{\mathfrak D}_\varphi + is}\right)
e^{i\pi\eta(s,{\mathfrak D}_\varphi)}.
\label{Z1}
\end{equation}
$Z(s,{\mathfrak D}_\varphi)$
satisfies the functional equation
$
Z(s,{\mathfrak D}_\varphi)Z(-s,{\mathfrak D}_\varphi)=
\exp \left(2\pi i \eta (s,{\mathfrak D}_\varphi)\right),
$
where the {\it twisted} zeta function $Z(s, \mD_\varphi)$ is meromorphic on $\mathbb C$.

It has been shown \cite{F1,F2} that for various flows the zeta function associated to any cyclic
flat bundle is actually meromorphic on a neighborhood of $[0, \infty)$, regular at $s=0$,
and its value at $s=0$ coincides with R-torsion with coefficients in the given flat bundle,
and thus is a topological invariant. Recall that Ray and Singer defined an analytic torsion
$\tau_\rho^{\bf an}(X)\in (0, \infty)$ for every closed Riemannian manifold $X$ and
orthogonal representation $\rho: \pi_1(X)\rightarrow O(n)$ \cite{RS}.
Because of the analogy with the Lefschetz fixed point formula, Fried proved that the geodesic
flow of a closed manifold of constant negative curvature has the Lefschetz property
\cite{F3}. He also conjectured that this remains true for any closed locally homogeneous
Riemannian manifold.

{\bf Zeta functions.} Zeta functions are given by the formulas \cite{Moscovici89}:
\begin{eqnarray}
Z(s, \mD) & = & \exp\left\{\!\!\sum_{[\gamma]\in {\cE_1(\Gamma)}}
\frac{(-1)^qL(\gamma, \mD)}
{\vert{\rm det}(I-P_h(\gamma))\vert^{1/2}}\frac{e^{-s\ell_\gamma}}{m_\gamma}\right\}\,,
\\
Z(s, \mD_\varphi) & = & \exp \left\{\!\!\sum_{[\gamma]\in {\cE_1(\Gamma)}}
{\rm Tr}\,\varphi (\gamma)\frac{(-1)^qL(\gamma, \mD)}
{\vert{\rm det}(I-P_h(\gamma))\vert^{1/2}}\frac{e^{-s\ell_\gamma}}{m_\gamma}\right\}\,,
\end{eqnarray}
where $\ell(\gamma)$ is the length of the closed geodesic $c_\gamma$ in the free homotopy
class corresponding to $[\gamma]$, $m(\gamma)$ is the multiplicity of $c_\gamma$,
$L(\gamma, \mD)$ are the Lefschitz numbers, and $P_h(\gamma)$ is the hyperbolic part of
the linear Poincar\'{e} map $P(\gamma)$ (see for detail \cite{Moscovici89}).

Fried's conjecture has been proved and an adequate theory of Selberg-type zeta functions
for locally symmetric spaces of higher rank was constructed in \cite{Moscovici91}.
Difficulties have been avoided by constructing certain {\it super} Selberg zeta functions,
$Z^\ell(s, \mD_\varphi)$, $0\leq \ell\leq 2m< {\rm dim}\,X$, as alternating products
of formal Selberg-like functions, which reduce to Selberg zeta functions only in the
three-dimensional rank one case. Each function $Z^\ell(s, \mD_\varphi)$ is
meromorphic on $\mathbb C$ and moreover satisfies a functional equation (see \cite{Moscovici91}
for details). Not surprisingly, the functional equations  play a crucial role in identifying
the special value of the Selberg-type spectral function ${\cR}(s; \varphi)$ with the R-torsion.
Finally,
${\cR}(s; \varphi)$ can be expressed as an alternating product of $Z$ (Cf. Eq. (\ref{Ruelle1})),
\begin{equation}
{\cR}(s; \varphi) = \prod_{\ell =0}^{{\rm dim}\,X -1} Z(s-{\rm dim}\,X+1+\ell ,
\,\mD_\varphi)^{(-1)^\ell}\,.
\end{equation}

\subsection{The Chern-Simons-- and \texorpdfstring{$\eta(s, \mathfrak D)$}{eta(s, D)}--invariants}
\label{Chern-Simons}

{\bf Gauge theory with finite gauge group.}
In the case of $n$-dimensional gauge theory with finite gauge group $G$ the path integral over a cosed
$n$-dimensional manifold $X$ reduces to a finite sum. This sum defines the quantum invarint $F(X)$.
We should note that for any manifold $M$ with ${\rm dim}\,M \leq n$, the groupoid of $G$-bundles
may be identified with the fundamental groupoid of the mapping space ${\rm Map}(M, BG)$, where
$BG$ is the classifying space carries a universal principal $G$-bundle \cite{Lurie}.
In the case of the 2-dimensional finite gauge theory based on the central extension
${\mathbb T}\rightarrow G^\tau\rightarrow G$ the classical invariant associated with a $G$-bundle
$P\rightarrow X$ over a closed oriented two-manifold, the {\it exponential action} of the field $\xi$
on $X$ is \cite{Lurie}
\begin{equation}
I(X, \xi) = e^{2\pi i \sigma_X(\xi)},
\label{action}
\end{equation}
where ${\mathbb T}= {\lambda\in {\mathbb C}: \vert\lambda\vert = 1}$ is the circle group,
$\sigma \in H^2(BG; {\mathbb Q}/{\mathbb Z})\cong H^3(BG; {\mathbb Z})$ is the characteristic class
corresponding to the central extension, and $\sigma_X(\xi)\in {\mathbb Q/{\mathbb Z}}$ is th
characteristic number of the bundle.
Then the quantum invariant of $X$ is
\begin{equation}
 F(X) = \sum_\xi\frac{1}{\#\ Aut\,\xi}I(X, \xi).
 \label{invariant}
\end{equation}
In Eq. (\ref{invariant}) the sum is over set of representative $G$-bundles on $X$, one in each
equivalence class. In the case of the trivial central extension $I(X, \xi) = 1$ for all $\xi$
and (\ref{invariant}) counts the number of representations of $\pi_1X$ into the finite group $G$.
In addition there is an overall factor of $1/\#\ G$ and one can extend to a theory of
unoriented manifolds.

{\bf Example: 2-dimensional theory over the circle $S^1$.}
The groupoid of $G$-bundles over $S^1$ is equivalebt to $G/\!/G$, where the group $G$ acts on
itself by conjugation (see for detail \cite{Lurie}). In the classical theory the value of
$I(S^1, \xi)$ is given by the central extension $G^\tau\rightarrow G$, viewed as an equivariant principal
$G$-bundle. It means that the value $I(S^1, \xi)$ at a bundle with holonomy $x\in G$ is the circle
torsor $G_x^\tau$ \cite{Lurie}. Finally in the quantum theory we take a sum over the appropriate
hermitian lines $K_\xi$ analogous to (\ref{invariant}). Thus we can compute $F(S^1)$ as the vector
space of central sections $K^\tau\rightarrow G$,
\begin{equation}
 F(S^1) = \bigoplus_\xi \frac{1}{\#\ Aut\,\xi}(K_\xi)^{Aut\, \xi}.
 \label{vs}
\end{equation}
Note that in expression (\ref{vs}) the metric on the hermitian line $K_x$ is scaled by the prefactor.
The quantum {\it path integrals} (\ref{invariant}) and (\ref{vs}) may be given in categorical language
\cite{Lurie}.

{\bf The Chern-Simons invariant.}
The $\eta$-invariant was introduced by Atiyah, Patodi, and Singer
\cite{APS} treating index theory on even dimensional manifolds with
boundary and it first appears there as a boundary correction in the usual local index formula.

Let as before $X$ be a closed odd dimensional spin manifold (which in their index theorem is
the boundary of an even dimensional spin manifold).
$\eta_{X}(s, \mD):= \eta(s, \mD_{\rho=trivial})$ is
analytic in $s$ and has a meromorphic continuation to $s\in {\mathbb C}$; it is regular
at $s=0$, and its value there is the $\eta$-invariant  of $\mD$ by setting
The result (\ref{Index}) holds for any Dirac operator on a $Spin^{\mathbb C}$ manifold coupled
to a vector bundle with connection (the metric of manifolds is supposed to be a product near
the boundary).

Let $X$ be a closed manifold, and $M$ be an oriented four-manifold with $\partial M=X$.
For a trivial representation $\rho$ one can choose a trivial
flat connection ${\underline A}$. Then for this choice we obtin
(see for detail \cite{BBG}):
\begin{equation}
{ Index}\,{\mathfrak D}_{\rho}  - {\rm dim}\,\rho \cdot { Index}\,{\mathfrak D}
= - CS({{\underline A}_{\rho}}) - \frac{1}{2}(\eta (0, {\mathfrak D}_\rho) -
{\rm dim} \rho \cdot \eta (0, {\mathfrak D})).
\label{Index2}
\end{equation}
The Chern-Simons invariant can be derived from Eq. (\ref{Index2}),
\begin{equation}
CS({\underline A}_{\rho})  = (1/2)({\rm dim} \rho \cdot \eta (0, {\mathfrak D})
- \eta (0, {\mathfrak D}_\rho))\,\, + \,\,
{\rm modulo} \,\,\,\, \mathbb Z \,.
\label{CS1}
\end{equation}

\begin{Resume}
\label{Resume}
A critical point of the Chern-Simons functional
is just a flat connection, and it corresponds to a representation of the fundamental
group $\pi_1(X)$. Thus, the value of this functional at a critical point can be regarded
as a topological invariant of a pair $(X, \rho)$, where $\rho$ is a representation of
$\pi_1(X)$. This is the Chern-Simons invariant of a flat connection on $X$.
We have derived Eq. {\rm (\ref{CS1})} for the Chern-Simons invariants of irreducible
$SU(n)$-flat connections on a locally symmetric manifolds of non-positive section curvature.
Taking into account that the Dirac operator is Hermitian, the function
$CS({\underline A}_{\rho})$ is real, and by making use Eq. {\rm (\ref{CS1})} we obtain
the exponential action for Chern-Simons theory in the form
\begin{equation}
\exp \left(2\pi i CS({\underline A}_{\rho})\right) = Z(0, {\mathfrak D})^{{\rm dim\,{\rho}}}\, Z(0, {\mathfrak D_\rho})
= (-1)^{{\rm dim}\,{\rho}\,{\rm dim}\,{\rm ker}\,{\mD}+ {\rm dim}\,{\rm ker}\,{\mD_\rho}}\,.
\label{ECS}
\end{equation}
There is an indeterminacy of sign unless
$
{\rm dim}\,{\rho}\,{\rm dim}\,{\rm ker}\,{\mD}+ {\rm dim}\,{\rm ker}\,{\mD_\rho} = 2n,
\,\, n\in {\mathbb Z}\,.
\label{sequential}
$
\end{Resume}

\subsection{Crossed products and deformation quantization}
\label{Crossed}

First we consider an interesting class of examples which is provided by the symmetric products
$S^nX = {X}^n/S_n$ of a space ${X}$. The following formula for the generating function of
the Euler characteristics
\begin{equation}
\sum_{n\geq 0}\chi({X}^n/S_n)q^n = \prod_{n=1}^\infty (1-q^n)^{-\chi({X})}.
\label{symmetric}
\end{equation}
has been proved in \cite{VW} by using the isomorphism
$\oplus_{n\geq 0}H^ \ast({X}^n/S_n)\cong S(\oplus_{n\geq 0}H^ \ast({X}))$.
Then we have
\begin{equation}
\prod_{n=1}^\infty (1-q^n)^{-\chi({X})} =
{\cR}(s = 1-i\varrho(\tau) )^{-\chi({X})}\,.
\end{equation}

{\bf Crossed products $\mathbb{C}[S_n]\ltimes \cA^{\otimes n}$.}
Recalling the cyclic homology and homology of an appropriate Lie algebras from Sect. \ref{cyclic}
we analyze the Hochschild homology of the crossed products
$\mathbb{C}[S_n]\ltimes \cA^{\otimes n}$ using the Hochschild homology
of the associative algebra $\cA$ (over $\mathbb C$).
Let us consider the Hochschild (co)homology of $\mathbb{C}[W]\ltimes \cA^{\otimes n}$. Here $\cA$
is the $q$-Weyl algebra or any its degeneration, $W$ is the Weyl group of type $A_{n-1}$ or $B_n$.

Let $X$ be an affine symplectic algebraic variety over $\mathbb C$. Suppose $\cA_+$ be a deformation quantization
of $X$. For $\cA = \cA_+[\hbar^{-1}]$ the Hochschild cohomology of the algebra $S^n\cA$, where the singular
Poisson variety $S^nX = X^n/S_n$, is additively isomorphic to the Chen-Ruan orbifold (or stringy)
cohomology of $S^nX$ with coefficients in ${\mathbb C}((\hbar))$ \cite{Etingof}.
Let $\cA(n):=\mathbb{C}[S_n]\ltimes \cA^{\otimes n}$. The generating function for ${\rm dim}HH^i(\cA(n))$
takes the form \cite{Etingof}:
\begin{eqnarray}
&& \sum t^iq^n\dim HH^i(\cA(n)) =\prod_{m\ge 1}\prod_{k\ge 0}
(1+(-1)^{k-1}q^mt^{k+d(m-1)})^{(-1)^{k-1}b_k(X)}
\nonumber \\
&& = \prod_{k\ge 0}\cR(s= (1+a+\varepsilon_1+\varepsilon_2)(1-i\varrho(\tau))-a+i\sigma(\tau))^{(-1)^{k-1}b_k(X)}.
\label{Hoch}
\end{eqnarray}
In Eq. (\ref{Hoch}) $b_k(X)$ are the Betti numbers of $\cA$, $a= d{\rm log}\,t/2\pi i\tau$,
$\varepsilon_1=(k-d){\rm log}\,t/2\pi i\tau$, and $\varepsilon_2= (k-1)/2\tau$.
Formulas for the generating series of the dimensions of the Hochschild cohomology of
$\mathbb{C}[W]\ltimes \cA^{\otimes n}$ may be given by analogy with  G\"{o}ttsche formula
(see for details \cite{Etingof}).

{\bf G\"{o}ttsche formula.}
For a one-dimensional higher variety (i.e. for a surface) the following results hold:
The generating function of the Poincar\'{e} polynomials $P_t(X^{[N]})$ of $X^{[N]}$ is given by
\begin{eqnarray}
\!\!\!\!\!&&\sum_{N=0}^{\infty}q^N\, P_t(X^{[N]})
 = \prod_{n=1}^\infty \frac{(1+t^{2n-1}q^n)^{b_1(X)}(1+t^{2n+1}q^n)^{b_3(X)}}
{(1-t^{2n-2}q^n)^{b_0(X)}(1-t^{2n}q^n)^{b_2(X)}(1-t^{2n+2}q^n)^{b_4(X)}}
\,.
\label{final}
\end{eqnarray}
As a result Eq.~\eqref{final} can be rewritten in terms of spectral Ruelle
functions \eqref{R1} and \eqref{R2}.

\section*{Acknowledgements}

We are much  grateful to Markku Oksanen for useful discussions and several improvements
in the manuscript.
AAB and AEG would like to acknowledge the Conselho Nacional
de Desenvolvimento Cient\'{i}fico e Tecnol\'{o}gico (CNPq, Brazil) and
Coordenac\~{a}o de Aperfei\c{c}amento de Pessoal de N\'{i}vel Superior
(CAPES, Brazil) for financial support.


\begin{thebibliography}{999999999}



\bibitem{Kac}
V. G. Kac, {\it Lie superalgebras}, Adv. in Math. {\bf 26} (1976) 8-96.

\bibitem{Davies}
B. Davies, O. Foda, M. Jimbo, T. Miwa, and A. Nakayashiki, {\it Diagonalization of the XXZ
Hamiltonian by vertex operators}, Commun. Math. Phys. {\bf 151} (1993) 89-154.

\bibitem{RS}
D. B. Ray and I. Singer, {\it R-Torsion and the Laplacian on Riemannian Manifolds},
Adv. Math. {\bf 7} (1971) 145-210.

\bibitem{Fried}
D. Fried, {\it Analytic torsion and closed geodesics on hyperbolic manifolds}, Invent. Math.
{\bf 84} (1986) 523-540.

\bibitem{Bytsenko3}
L. Bonora and A. A. Bytsenko, {\it Partition Functions for Quantum Gravity, Black Holes, Elliptic Genera
and Lie Algebra Homologies}, Nucl. Phys. B {\bf 852} (2011) 508-537.

\bibitem{Bytsenko4}
L. Bonora, A. A. Bytsenko and E. Elizalde, {\it String partition functions, Hilbert schemes and affine
Lie algebra representations on homology groups}, J. Phys. A: Math. Theor. {\bf 45} (2012)
374002 (41 pp).

\bibitem{Kac_book}
V. G. Kac, {\it Infinite dimensional Lie algebras}, Cambridge University Press 1990.


\bibitem{Perry}
P. Perry, {\it A Poisson summation formula and lower bounds for resonances in hyperbolic
manifolds}, Int. Math. Res. Notes {\bf 34} (2003) 18371851.

\bibitem{PW}
P. Perry and F. Williams, {\it Selberg zeta function and trace formula for the BTZ
black hole}, J. of Pure and Applied Math. {\bf 9} (2003) 1-21.

\bibitem{Gottsche}
L. G\"{o}ttsche, {\it The Betti numbers of the Hilbert Scheme of Points on a Smooth Projective Surface},
Math. Ann. {\bf 286} (1990) 193-207.

\bibitem{Kac2}
V. Kac and D. Peterson, {\it Infinite dimensional Lie algebras, theta functions and modular
forms}, Advances in Math. {\bf 53} (1984) 125-264.

\bibitem{Belavin1}
A. A. Belavin, A. M. Polyakov and A. B. Zamolodchikov, {\it
Infinite conformal symmetry of critical fluctuations in two
dimensions}, J. Stat. Phys. {\bf 34} (1984) 763-774.

\bibitem{Belavin2}
A. A. Belavin, A. M. Polyakov and A. B. Zamolodchikov, {\it
Infinite conformal symmetry in two-dimensional quantum field
theory}, Nucl. Phys. B {\bf 241} (1984) 333-380.

\bibitem{Dowker}
J. S. Dowker, {\it Modular properties of Eisenstein series and
statistical physics}, arXiv:0810.0537 [hep-th].

\bibitem{Fuks}
D. B. Fuks, {\it Cohomology of Infinite-Dimensional Lie Algebras}, Contemporary Soviet Mathematics,
Consultas Bureau, New York, 1986.

\bibitem{Troitsky}
Yu. P. Soloviev and E.V. Troitsky, {\it $C^\ast$-algebras and elliptic operators in differential
topology}, Translations of Mathematical Monographs, American Mathematical Society,
Providence, Rhode Island {\bf 192}.

\bibitem{Moscovici}
H. Moscovici, {\it Cyclic Cohomology and Invariants of Multiply Connected Manifolds},
Proceedings of the International Congress of Mathematics, Kyoto, Japan, 1990,
{\bf 1} (1991) 675-688.

\bibitem{APS}
M. Atiyah, V. K. Patodi and I. M. Singer, {\it Spectral asymmetry and Riemannian geometry I.},
Proc. Camb. Phil. Soc. {\bf 77} (1975) 43--69;
Part II, ibid. {\bf 78} (1975) 405--432;
Part III, ibid. {\bf 79} (1976) 71--99.


\bibitem{RS2}
D. B. Ray and I. Singer, {\it Analytic torsion for complex manifolds}, Ann. Math.
{\bf 98} (1973) 154-177.

\bibitem{Karoubi}
M. Karoubi, {\it K-Theory. An Introduction}, Springer-Verlag, Berlin Heidelberg New York,
1978.

\bibitem{FF}
A. T. Fomenko and D. B. Fuks, {\it Homotopical Topology (Graduate Texts in Mathematics)},
Nauka, Moscow, 1989 (edition in Russian), Springer.

\bibitem{Witten85}
E. Witten, {\it Global gravitational anomalies}, Commun. Math. Phys. {\bf 100} (1985)
197-229.

\bibitem{Cayley}
A. Cayley, {\it About the algebraic structure of the orthogonal group and the other classical
groups in a field of characteristic zero or a prime characteristic}, J. Reine Angew. Math.
{\bf 32} (1846) 119-123.

\bibitem{Moscovici89}
 H. Moscovici and R. Stanton, {\it Eta invariants of Dirac operators on locally symmetric
manifolds}, Invent. Math. {\bf 95} (1989) 629-666.

\bibitem{Nishi}
H. Nishi, {\it $SU(n)$-Chern-Simons Invariants of Seifert Fibered 3-Manifolds},
International Journal of Mathematics {\bf 9} (1998) 295-330.

\bibitem{Moscovici91}
H. Moscovici and R. Stanton, {\it R-torsion and zeta functions for locally symmetric
manifolds}, Invent. Math. {\bf 105} (1991) 185-216.

\bibitem{F1}
D. Fried, {\it Lefschetz formulas for flows}, in the Lefschetz Centennial Conference,
part III, Contemporary Mathematics, {\bf 58} (1987) 19-69.

\bibitem{F2}
D. Fried, {\it Counting circles}, Dynamical Systems, Lecture Notes in Mathematics,
{\bf 1348} (2006) 196-215, Springer-Verlag.

\bibitem{F3}
D. Fried, {\it Analytic torsion and closed geodesic on hyperbolic manifolds},
Invent. Math. {\bf 84} (1986) 523-540.

\bibitem{Lurie}
D. S. Freed, M. J. Hopkins, J. Lurie and C. Teleman, {\it Topological
quantum field theories from compact Lie groups}, arXiv:math.AT/0905.0731v2 [math.AT],
in proceedings of Conference: C08-06-09.7 (May 2009, 39 pp).

\bibitem{BBG}
L. Bonora, A. A. Bytsenko and A. E. Gon\c{c}alves, {\it Chern-Simons Invariants on Hyperbolic Manifolds and
Topological Quantum Field Theories}, Eur. Phys. J. C (2016) 76:625; arXiv:1606.02554v1
[hep-th].

\bibitem{VW}
C. Vafa and E. Witten, {\it A strong coupling test of S-dualitty}, Nucl. Phys. B {\bf 431}
(1994) 3-77.

\bibitem{Etingof}
P. Etingof and A. Oblomkov, {\it Quantization, Orbifold Cohomology, and Cherednik Algebras},
arXiv:0311005v5 [math.QA], Macdonald 75th birthday volume.

\end{thebibliography}
\end{document}